\documentclass[aps,amsfonts,pra,twocolumn,showpacs]{revtex4-1}
\usepackage{epsfig,amsmath,amssymb,bm,epsf,graphicx,psfrag}
\usepackage[all]{xy}
\usepackage{color}

\def\ket#1{\vert#1\rangle}
\def\ketbra#1{\vert#1\rangle\langle#1\vert}

\def\Longarrow{\protect\@lra}
\def\@lra{\relbar\joinrel\relbar\joinrel\relbar\joinrel%
          \relbar\joinrel\rightarrow}

\newcommand{\bc}{\begin{center}}
\newcommand{\ec}{\end{center}}
\newcommand{\be}{\begin{equation}}
\newcommand{\ee}{\end{equation}}
\newcommand{\bea}{\begin{eqnarray}}
\newcommand{\eea}{\end{eqnarray}}

\newcommand{\ncd}{\newcommand}
\ncd{\QCcns}{$QC_{\cal{C}}$}
\ncd{\QCc}{$QC_{\cal{C}}\;$}

\newtheorem{Lemma}{Lemma}

\definecolor{libl}{cmyk}{0.2,0.1,0,0}

\def\dK{{{\rm dim}\big({\rm ker}(H)\big)}}
\begin{document}

%\title{The spin-2 Affleck-Kennedy-Lieb-Tasaki state on the square lattice is a universal resource for quantum computation}
\title{Universal measurement-based quantum computation with spin-2 Affleck-Kennedy-Lieb-Tasaki states}
\author{Tzu-Chieh Wei}
\affiliation{C. N. Yang Institute for Theoretical Physics and
Department of Physics and Astronomy, State University of New York at
Stony Brook, Stony Brook, NY 11794-3840, USA}
\author{Robert Raussendorf}
\affiliation{Department of Physics and Astronomy, University of
British Columbia, Vancouver, British Columbia, V6T 1Z1, Canada}
\begin{abstract}
We demonstrate that the spin-2 Affleck-Kennedy-Lieb-Tasaki (AKLT)
state on the square lattice is a universal resource for the
measurement-based quantum computation. Our proof is done by locally
converting the AKLT to two-dimensional random planar graph states and by certifying that
with high probability the resulting random graphs are in the
supercritical phase of percolation using Monte Carlo simulations. One key enabling point is the exact weight formula that we derive for arbitrary measurement outcomes according to a spin-2 POVM on all spins. We also argue that the spin-2 AKLT state on the three-dimensional diamond lattice is a universal resource, the advantage of which would be the possibility of implementing fault-tolerant quantum computation with topological protection. In addition, as we deform the AKLT Hamiltonian, there is a finite region that the ground state can still support a universal resource before making a transition in its quantum computational power.
\end{abstract}
\date \today
 \maketitle

 \section{Introduction and motivation}

Quantum computation (QC) can be implemented in various 
frameworks, such as the standard circuit model~\cite{NielsenChuang}, adiabatic evolution~\cite{Farhi},
manipulation of exotic anyons in topological phases, and local
measurement on certain entangled states~\cite{TQC}. In
the measurement-based model of quantum computation (MBQC)~\cite{Oneway,Oneway2,RaussendorfWei12}, only certain entangled states are
known to  provide the capability for driving a universal quantum
computation via local measurement, such as the cluster state on the square lattice~\cite{cluster}. A complete classification of entanglement structure that enables
MBQC remains a challenging open question. Moreover, whether these entangled states arise as unique ground states of short-range gapped Hamiltonians is relevant for robust resource state creation and possibly further protection during computation~\cite{Nielsen}. From this latter viewpoint, cluster states, unfortunately, cannot be the unique ground state of two-body interacting qubit Hamiltonains~\cite{Nielsen,NoGo}, albeit they can be approximately~\cite{approx}. The key obstacle for complete characterization is that there is no simple physical observables (or order-parameter-like quantities) that is generic for answering whether a state is universal or not. Proving either universality or non-universality is thus in general highly nontrivial. In order to make progress toward complete characterization of universal resource states, it requires a substantial breakthrough in understanding the entanglement structure necessary for realizing universal gates.

Until now the most complete characterization is for 1D resource states, even though they are not universal for QC. This includes 1D cluster state, 1D spin-1 AKLT state, and matrix-product states of certain yet general forms~\cite{Gross,GrossEtAl,GrossCompWeb,Chen2010}.
However, 2D and higher dimensions are much less understood. After the disovery of the cluster state on the square lattice,  it was recognized that the generalization of the cluster state---the graph state---also provides universal resource on various other 2D regular lattices~\cite{Universal}. Furthermore,  it was also shown how to characterize quantum computational universality for graph states on faulty lattices~\cite{Browne} as well random 2D planar graphs~\cite{WeiAffleckRaussendorf12}.  The issue of the universality in the family of cluster or graph states is well understood due to their simple entanglement structure. Beyond this family of states, it seems only a handful of other entangled states are known to be universal~\cite{Gross,Chen,Cai10,ZengKwek}. Deciding whether a given quantum state is a universal resource for MBQC is still a challenge, let alone generalization to a family of states. No generally applicable strategies to answer this question are known.

 An interesting and emerging family that may potentially be as useful with regards to resourcefulness for MBQC is the so-called Affleck-Kennedy-Lieb-Tasaki
 (AKLT) states~\cite{AKLT}.  Originally, the 1D spin-1 AKLT chain provides the first evidence to support Haldane's conjecture on the spectral gap property of integer-spin chains with rotational symmetry~\cite{Haldane}. It also gives the first instance of the matrix product states~\cite{MPS}, and more recently serves as an example of a symmetry-protected topological ordered state~\cite{Wen}. The extension to two dimensions, such as the ones on the honeycomb and square lattices, provides illustration of projected entangled pairs states~\cite{PEPS} and models of spin rotational symmetry with potentially a finite gap above the ground state~\cite{AKLT}. 
In three dimensions, some AKLT states can become N\'eel ordered~\cite{Param}. Similar to graph states, AKLT states can be defined
 on any graph in terms of the valence-bond picture and local symmetrization, but the local spins can be of magnitude 1, 3/2, 2 or higher.  Moreover, they are unique ground states of two-body interacting Hamiltonians with suitable
 boundary conditions.
 Even though quantum computational universality in MBQC requires at least a 2D structure, 
 the results on the 1D spin-1 AKLT state for theoretically and experimentally simulating
 one-qubit gates~\cite{Gross,Brennen,Resch} prompted the quest of universality in 2D AKLT states.
 However, the capability of full quantum computational universality was only established
 recently in the spin-3/2 AKLT state on the honeycome lattice~\cite{WeiAffleckRaussendorf11,Miyake11} and later on some other trivalent
 lattices~\cite{Wei13},
as well as some lattices that host spin-2 and other lower spin (such
 as spin-3/2 or spin-1)  hybrid AKLT states ~\cite{WeiEtAl}. No AKLT states of uniform  spin-2 entities have been known to provide universal resource, and on the contrary, the AKLT state on the kagom\'e lattice was argued  to be non-universal.

Here we demonstrate that  the spin-2 AKLT
 state on the square lattice is indeed a
universal resource for MBQC. 
 This result adds a missing piece to a series of study~\cite{WeiAffleckRaussendorf11,Miyake11,WeiAffleckRaussendorf12,Wei13,WeiEtAl} and gives 
rise to the following emerging picture that advances our understanding of the quantum computational universality in the valence-bond family.  AKLT states involving
 spin-2 and other lower spin entities are universal if they reside on
 a two-dimensional frustration-free regular lattice with any
 combination of spin-2, spin-3/2, spin-1 and spin-1/2 (consistent
 with the lattice). Furthermore,  geometric frustration can, but not necessarily, be a hinderance to the quantum computational universality, and a frustrated lattice can be decorated (by adding additional spins) such that the resultant AKLT state is universal. Additionally, we argue that the spin-2 AKLT state on the three-dimensional diamond lattice is also a universal resource. The advantage of using a three-dimensonal resource state would be the possibility of implementing fault-tolerant quantum computation with topological protection~\cite{TO}.

The family of AKLT states also provides the basis for
generalization. They can be deformed so as to maintain universality in a range of the deformation parameter~\cite{DarmawanBrennenBartlett}. We shall also argue that there is a finite region around the spin-2 AKLT point such that the ground state of a deformed AKLT Hamiltonian still supports a universal resource; see below in Sec.~\ref{sec:Deformed}. Furthermore, frustrated AKLT states that are not likely universal~\cite{Wei13} can be deformed in a such way that they are connected continuously to a cluster state~\cite{Darmawan2}.  These can be used to study the connection
of transitions in phases of matter and in quantum computational power~\cite{GrossEtAl,thermal,DarmawanBrennenBartlett,WeiLiKwek}.    With suitably chosen boundary conditions, AKLT states are unique ground states of certain two-body interacting Hamiltonians~\cite{AKLT}, some of which are believed to possess finite spectral gap~\cite{Ganesh,Garcia}, including the spin-3/2 and the spin-2 AKLT states on the honeycomb and square lattices, respectively. A finite gap and the uniqueness of the ground state is a desirable feature for creating the resource state by cooling the system~\cite{Nielsen,Darmawan2,CoolCluster,DemonCooling}.

The remainder of this paper is organized as follows.  In Sec.~\ref{sec:strategy} we describe the overall strategy for showing that an AKLT state is a universal resource for quantum computation.  It consists of two steps, namely the mapping of the AKLT state to a random planar graph state by applying a suitable POVM, and the numerical demonstration that, in the typical case, the resulting graph state can be mapped to a two-­dimensional cluster state by further local measurements.  These two steps are described in detail in Sections~\ref{Reduc} and~\ref{es}, respectively.  In Sec.~\ref{sec:Deformed} we discuss one possible extension of our techniques to transitions in quantum computational power away from the AKLT point. We conclude in Section~\ref{Sum} and also discuss possible future directions and potential experimental realizations.

\section{Overall strategy}
\label{sec:strategy}
Our goal is to show that any quantum computation that is efficiently implemented in the circuit model can also be implemented efficiently by a sequence of adaptive local measurements on a spin-2 AKLT state. In other words, we want to show that the spin-2 AKLT state is a universal resource for MBQC.

The overall strategy for demonstrating this is the so-called quantum state reduction~\cite{Chen2010}, i.e., to show that, by local measurement, the state can be converted to a known resource state, such as cluster states, with finite probability, even in the limit of large system sizes. In our previous study of the spin-3/2 AKLT state~\cite{WeiAffleckRaussendorf11}, the reduction proceeds in three steps:
\begin{enumerate}
\item{Devise a pattern of local measurements that transforms the AKLT state into a graph state $|G\rangle$, where the graph $G$ depends on the random measurement outcomes but is always planar. This proceeds in two sub-steps, namely (1a) the creation of an encoded graph state $\overline{|G\rangle}$ by local generalized measurements, and (1b) a decoding of this graph state by local projective measurements. The support of each encoded qubit in $\overline{|G\rangle}$ after step (1a) is called a ``domain'' of the AKLT spin lattice ${\cal{L}}$. The domains fluctuate in size.}
\item{Show that a planar graph state $|G\rangle$ can be reduced to a 2D cluster state (the standard universal resource), if the domains are all small and traversing paths through $G$ exist.}
\item{Numerically demonstrate that, for typical POVM outcomes in Step 1, the graph states produced  satisfy the pre-conditions of Step 2.}
\end{enumerate}

It is not {\it a priori\/} obvious that such procedure can work for the spin-2 or higher-spin AKLT states. Indeed it remains open whether one can even find a suitable POVM  for AKLT states with spin magnitude higher than two to reduce them to graph states. For the spin-2 case, 
the straightforward generalization of the POVM used in the case of spin 3/2 is no longer a POVM for spin 2. This is overcome by adding local POVM elements~\cite{WeiEtAl} such that the resulting measurement still maps to an encoded graph state, and the encoding can still be undone by local projective measurements. The additional local POVM elements amount to further projective measurement and hence disentanglement of the spins from the remaining.   However, the new POVM does not guarantee that the resulting graph state $|G\rangle$  corresponds to a {\em{planar}} graph $G$, but the planarity of $G$ is a requirement for Step 2 to work. Fortunately, the obstructions to planarity are local, and we can append a further round of  measurements to remove them, thereby restoring planarity at the cost of reducing connectivity. We refer to this latter procedure as {\it restoration of planarity by thinning\/}; see Sec.~\ref{sec:thinning}. 

Once we obtain planar graph states,  Step 2 of reduction to 2D cluster states goes through unchanged for the spin-2 case.  2D Cluster states are the standard universal resource for MBQC, i.e., further local measurements on such a state can then implement any desired quantum circuit~\cite{Oneway,Oneway2,RaussendorfWei12}. 

Regarding Step 3, for correct numerical simulation of the measurement procedure, we require  an efficient method for calculating the exact probability weights of the POVM outcomes. The probabilities for the outcomes $F_\alpha$ and $K_\alpha$ on each individual site are 4/15 and 1/15, respectively. Beyond those values, there are higher-order correlations between POVM outcomes on different sites which we cannot simply neglect. Even more seriously, some randomly assigned $\{F,K\}$ do not occur, i.e., their probability is exactly zero. How do we know what outcomes can occur and with what probabilitity? It turns out that there is a closed-form expression for the exact probability weights which can be efficiently evaluated; see Sec.~\ref{sec:weight} for the expression and Appendix~\ref{sec:proof2} for the proof.

The most pronounced difference between the spin-3/2 and spin-2 probability weights is that for spin 3/2 all possible combinations of POVM outcomes do indeed occur with non-zero probability (except when the lattice is not bi-colorable, i.e., due to geometric frustration). This arises a consequence of the bi-colorability of the underlying honeycomb lattice. For the spin-2 case, as already mentioned, certain combinations of POVM outcomes do not occur, i.e., have probability zero. The underlying spin lattice ${\cal{L}}$ (a square lattice) is still bi-colorable but this is no longer the deciding factor.\medskip

Let us summarize the procedure to establish the universality of the spin-2 AKLT state on the square lattice. The generalization of it to the diamond lattice is straightfoward but will not be carried out here. 

\smallskip\noindent {\bf Procedure for establishing universality}:
\begin{enumerate}
 \item{ Use the weight formula in Lemma 1 to sample typical POVM outcomes. }
 \item{Apply the thinning proceudre in Sec.~\ref{sec:thinning} to obtain assoicated random planar graphs. (One should not confuse the thinning procedure with the deletion used in checking the robustness of the graph connectivity, presented in the next step.) }
 \item{ Check whether there is a traversing path and record the probability $p_{\rm span}$ that this occurs. If so, examine how robust the connectivity in the graphs. To do this, we employ the idea from percolation and delete every vertex with a probability $p_{\rm delete}$ and record the probability $p_{\rm span}$ that a traversing path still exists. }
 \end{enumerate}
If we can demonstrate, from the behavior of $p_{\rm span}$ vs. $p_{\rm delete}$ for different sizes $L$, that there is a phase transition (say at $p_{\rm delete}^*$), then according to the theory of percolation, the graphs that reside in the phase with $p_{\rm delete} < p_{\rm delete}*$ (a.k.a. the supercritical phase) contain macroscopic number of traversing paths. As we shall demonstrate in Sec.~\ref{sec:numerics} that this is indeed the case, and hence the random graph states are universal, implying the original AKLT state is also universal. 
\begin{figure}
 \includegraphics[width=0.5\textwidth]{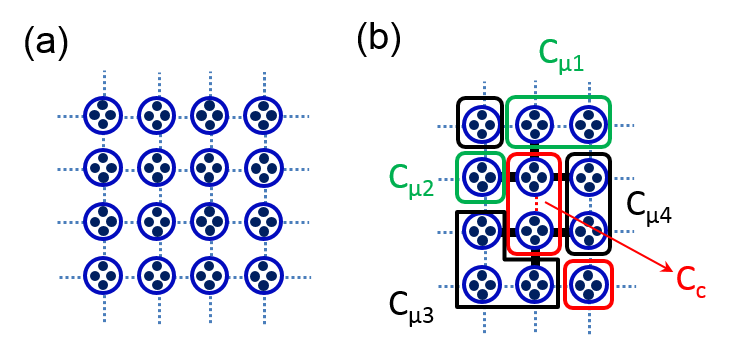} 
        \caption{ (a)  AKLT state on the square lattice. Spin singlets $|\phi\rangle_e=(|01\rangle-|10\rangle)/\sqrt{2}$ of two virtual spins 1/2 are located
  on the edges of the square lattice, indicated by dashed lines. A projection at each lattice site onto the symmetric subspace of four virtual spins creates the AKLT state.  (b) Illustration of domains and virtual qubits inside domains. Five of the domains are labeled: ${\cal C}_C, {\cal C}_{\mu1}, {\cal C}_{\mu2}, {\cal C}_{\mu3}, {\cal C}_{\mu4}$. 
         \label{fig:square} }
\end{figure}

\section{Reduction from AKLT states to graph states}
\label{Reduc}
Let us define the AKLT state on the square lattice. It is useful to view the spin-2 particle on each site is consisting of four virtual qubits. Each virtual qubit forms a singlet state, $|\phi\rangle_e=(|01\rangle-|10\rangle)/\sqrt{2}$, with its corresponding virtual qubit on the neighboring site, with the singlets indicated by the dotted edges; see Fig.~\ref{fig:square}. In order to yield a local five-level spin-2 particle, a local projection on each site is made from the Hilbert space of four virtual qubits to their symmetric subspace, which is isomorphic to the spin-2 Hilbert space~\cite{AKLT}.

\subsection{Reduction from spin-2 entities to qubits: the generalized measurement}
\label{sec:reduction}
The POVM  we shall employ consists of three rank-two
elements and three additional rank-one elements~\cite{WeiEtAl}:
\begin{subequations}
\label{POVMspin2}
\begin{eqnarray}
\!\!\!\!\!F_\alpha&=&\sqrt{\frac{2}{3}}\,(\ketbra{{S_\alpha\!=\!2}}+\ketbra{{S_\alpha\!=\!-2}})\\
\!\!\!\!\!K_\alpha&=&\sqrt{\frac{1}{3}}\,\ketbra{\phi_\alpha^-},
\end{eqnarray}
\end{subequations} where
 $\alpha=x,y,z$ and
$\ket{\phi_\alpha^\pm}\equiv(\ket{{S_\alpha\!=\!2}}\pm\ket{{S_\alpha\!=\!-2}})/\sqrt{2}$. The $F$'s are straightforward generalization from the spin-3/2 case~\cite{WeiAffleckRaussendorf11}, but they do not give rise to the completeness relation, which is required for conservation of probabilities. By adding $K$'s,
it can be verified that the completeness relation is satisfied:
$\sum_\alpha {F}^\dagger_\alpha {F}_\alpha +\sum_\alpha
{K}^\dagger_\alpha {K}_\alpha=\openone$, i.e., proving that there only six possible outcomes associated to  ${F}_{\alpha=x,y,z}$ and $K_{\alpha=x,y,z}$. A spin-2 particle in state $|\Phi\rangle$ that undergoes such a generalized measurement becomes either $F_\alpha|\Phi\rangle$ or $K_\alpha|\Phi\rangle$.
  
The reduced density matrix for
a single site of the AKLT state is a completely mixed state, and
therefore, each unwanted type occurs on average with a probability
$1/15$. We note that the outcomes of K result in projection to a one-dimensional subspace (instead of a two-dimensional subspace that could form the basis of a qubit), they are thus  regarded as ``undesired'' or ``unwanted'' outcomes. An unwanted outcome associated with $K$ {thus} occurs with
probability $p_{\rm err}=3\times1/15=1/5$. However, as we shall see below in Sec.~\ref{sec:weight} that not all POVM outcomes associated with sets of $\{F_{\alpha(v)},K_{\beta(w)}\}$  occur with non-zero probability, due to the correlation present in the AKLT state. Below we discuss the effect of $F$ and $K$ outcomes.

For bi-colorable graphs, such as the square lattice, any all-$F$ POVM outcome can occur.
We have previously shown that in this case, the post-measurement state
\begin{equation}
\label{eqn:G0}
\overline{|G_0(\{F\})\rangle}\sim\bigotimes_{v\in {\cal L}} {F}_{\alpha(v)}|\psi_{\rm
AKLT}\rangle
\end{equation}
is an encoded graph
state~\cite{WeiAffleckRaussendorf11,WeiAffleckRaussendorf12}, whose properties are described below.

\begin{enumerate}
\item{Each domain on ${\cal{L}}$ supports a single encoded qubit, i.e., the domains $D \subset {\cal{L}}$ are the sites or vertices of the graph $G_0$, with the encoding as described in Table~\ref{tbl:encoding}. The encoded qubits form a graph state $\overline{|G_0\rangle}$. When there is no confusion, we shall not distinguish between the graph state $|G_0\rangle$ and its encoded version $\overline{|G_0\rangle}$ and omit the labeling $\{F\}$.}
\item{The graph $G_0$ has an edge between the vertices $v(D)$ and $v(D')$, if the domains $D$ and $D'$ are connected by an odd number of edges in ${\cal{L}}$.} 
\item{Be $D$ a domain of type $T\in \{x,y,z\}$ with $n_\alpha$ neighbouring domains of type $\alpha$. The stabilizer operators for such a graph state are shown in Eq.~(\ref{eqn:stablizer}) in terms of encoded logical operators. They are characterized by the so-called stabilizer matrix, and in the case of graph state, is given via the adjacency matrix $A_{G_0}$ of the graph $G_0$. It is seen that when 
$$
\begin{array}{rl}
n_y \mod 2 =1,& \text{for }T=x,\\
n_x \mod 2 =1, & \text{for }T=y,\\
n_y \mod 2 =1, & \text{for }T=z,
\end{array}
$$
the stabilizer operator ${\cal K}_D$ has a logical $Y$ operator at the support of $D$. This means that the  graph $G_0$ has a self-loop attached to the domain $D$, i.e., $\left(A_{G_0}\right)_{D,D}=1$. }
\end{enumerate}

We recall the definition of a ``domain''. A domain is a maximal set of neighbouring sites in the lattice ${\cal{L}}$ for which the outcome of the POVM Eq.~(\ref{POVMspin2}) is $F_\alpha$ or $K_\alpha$ with the same $\alpha$. That is, there are domains of $x$, $y$ and $z$-type, and neighbouring domains must be of different type. The self-loop is a convenient picture to visualize the graph. But we can perform local logical rotation to transform $Y$ to $X$ so as to remove the self-loop, then the resulting stabilizer operators will be in the canonical form. (Such rotation will also change the basis of logical measurement.) Moreover, we shall often not distinguish between an encoded $\overline{X}$ or $\overline{Y}$ operator from the corresponding $X$ or $Y$ operator, unless necessary. We also note that the stabilizer operator for graph states is usually defined as
${\cal K}_D\equiv \pm X_D \bigotimes_{D'\in {\rm nb}(D)} Z_{D'}$, where ${\rm nb}(D)$ denotes the set of $D$'s neighbors, i.e., those vertices connected to $D$ by an edge. We shall refer to such stabilizer operators as being in the canonical form and we have allowed additional signs in the definition. The graph state $|G\rangle$ can thus be defined by ${\cal K}_D|G\rangle=|G\rangle$ for all vertices $D$. However, under a local basis change, the stabilizer operator can be transformed to the form ${\cal K}_D\equiv \pm Y_D \bigotimes_{D'\in {\rm nb}(D)} Z_{D'}$, where the operator at $D$ is the logical $Y$. The above point 3 is to determine exactly what form of the stabilizer is for each domain belonging to the graph state $|G_0\rangle$; see also Eq.~(\ref{eqn:stablizer}).

We discuss the effect of $K$'s in the next subsection. 
But let us remark that the effect of the POVM~(\ref{POVMspin2}) is to produce an encoded graph state $\overline{|G\rangle}$ corresponding to a graph $G$ with adjacency matrix $A_G$. In the previous case of spin 3/2 it was sufficient to identify the graph state only up to local unitary equivalence. For the spin-2 case, due to the more complicated POVM and weight formula, this is no longer the case. In particular, we need to keep track of the self-loops in $G$ and the eigenvalues of the stabilizer generators for $\overline{|G\rangle}$. It is useful to use the graph state $|G_0\rangle$ as a reference point for subsequent discussions. 

\subsection{POVM  outcomes $K_\alpha$: domain shrinking and logical Pauli measurements} 
\label{sec:K}
We now discuss the effect of $K$'s, which can be rewritten as  follows,
\begin{equation}\label{KeKF}
K_\alpha = \sqrt{1/2} \ketbra{\phi_\alpha^-}
{F}_\alpha=\sqrt{2/3}\, K_\alpha F_\alpha.
\end{equation}
We can thus think of the POVM Eq.~(\ref{POVMspin2}) as a two-stage process: first the outcomes on all sites are $F$'s, and then a number of sites are flipped to $K$ or equivalently a projective measurement is done in the basis $\ket{\phi_\alpha^\pm}$ and the result $\ket{\phi_\alpha^-}$ is post-selected.

We shall denote  by $\{F,K\}$  the POVM outcomes on all sites,  by $J_F
\subset {\cal L}$ the set of sites where the POVM outcome is of
$F$-type, and by $J_K={\cal L}\backslash J_F$ the set of sites where POVM
outcome is of $K$-type. 
Upon obtaining $\{F,K\}$ we can deduce the state $|G\rangle$ that the original AKLT state is transformed to,
\begin{eqnarray*}
&&|G(\{F,K\})\rangle= \bigotimes_{u\in J_K} K_{\alpha(u)} \bigotimes_{v\in J_F} F_{\alpha(v)} |\psi_{\rm AKLT}\rangle\\
&&= \left(\sqrt{\frac{1}{2}}\right)^{|J_K|} \bigotimes_{u\in J_K} \ketbra{\phi_{\alpha(u)}^-}  \bigotimes_{v\in {\cal L}} F_{\alpha(v)} |\psi_{\rm AKLT}\rangle,
\end{eqnarray*}
where the state is not normalized and the probability of the set of POVM outcomes $\{F,K\}$ occurs is 
\begin{equation}
p(\{F,K\})=\langle G(\{F,K\})|G(\{F,K\})\rangle.
\end{equation}
It was shown previously~\cite{WeiAffleckRaussendorf11} that
\begin{equation}
\label{eqn:Fweight} \bigotimes_{u\in {\cal L}}
F_{\alpha(u)} |{\psi}_{\rm AKLT}\rangle=c_0\, \left(\frac{1}{\sqrt{2}}\right)^{|{\cal
E}|-|V|}|G_0\rangle,
\end{equation}
where  $c_0$ is an outcome-independent overall normalization, $V$ is the set of domains, ${\cal E}$ is the set of
inter-domain edges (before the modulo-2 operation) and $|G_0\rangle$ is properly normalized to have unit norm~\cite{WeiAffleckRaussendorf11}.  For the encoding using virtual-qubit picture, see Table~\ref{tbl:encoding}.

Summarizing the above discussion, we have
\begin{eqnarray}
&&|G(\{F,K\})\rangle
= c_0\,\left(\sqrt{\frac{1}{2}}\right)^{|{\cal E}|-|V|+|J_K|} \nonumber\\
&&\quad\left(\bigotimes_{u\in J_K} \ketbra{\phi_{\alpha(u)}^-}\right) |G_0(\{F\})\rangle,
\end{eqnarray}
where $|G_0(\{F\})\rangle$ is assumed to be properly normalized.
Without the additional operators $\bigotimes_{u\in J_K} \ketbra{\phi_{\alpha(u)}^-}$ the analysis of the computational universality would be the same as in the spin-3/2 case. It is these operators that complicate the situtation. However, as we shall see below their effect is not serious.

First, as an example, consider a $z$-domain with two sites, one of which is measured in $F_z$ and the other in $K_z$. As Table~\ref{tbl:encoding} shows, the effect of the measurement in $K_z$ is a mere shrinking of the domain from two sites to one. The graph $G$ underlying the encoded graph state $\overline{|G\rangle}$ remains the same, only the encoding changes on the domain in question. Strictly speaking, the normalization is reduced by a factor of $\sqrt{2}$, due to the post-selection of only the `-' outcome.

As a second example, consider a $z$-domain comprising a single site and this site is affected by a $K_z$. In this case, the effect of the POVM element $K_z$ on $\overline{|G_0\rangle}$ is different: the encoded qubit living on that domain is projected into an eigenstate of $\overline{X}$ (with eigenvalue $-1$).   In general  the effect of all  $\ketbra{\phi_\alpha^-}$ (associated with $K_\alpha$) in multi-site domains ${\cal C}$, including the correct normalization factor,  is equivalent to  a logical measurement of $X$ operator, 
\begin{equation}
\label{eqn:Pc}
P_c=[1+O_c]/2^{|V_c|}=[1+(-1)^{|V_c|} X_c]/2^{|V_c|},
\end{equation}
where $|V_c|$ denotes the number of sites in the domain and we have defined $O_c\equiv (-1)^{|V_c|} X_c$;  see below in
Appendix~\ref{sec:VAstablizer}.  (In that section we also derive the form of the stabilizer for any domain, and it is seen that it is not necessary in the canonical graph-state form with $X$ at a given vertex and $Z$'s at neighboring sites.) In the canoical graph-state basis (CGSB) this measurement corresponds to either the logical $X$ or $Y$ basis. If there is no self-loop on this domain, then the measurement in the CGSB is in the logical $X$ basis and we shall refer to this domain as an $X$-measured domain. If there is a self-loop on this domain, we shall refer to this domain as a $Y$-measured domain as the measurement in the CGSB is in the logical $Y$ basis, for which the graphical rule is to perform local complementation before removing the vertex  (see below in
Sec.~\ref{sec:VAstablizer}). 
% In the canonical graph-state picture, where we rotate locally so that ${\cal K_C}|_C = X_C$, the measurement will be $Y$ operator (which is the effect of the self-loop). 

 In fact the above two examples give the complete account of the effects caused by the POVM outcomes $K_\alpha$. We need to discuss each domain separately, and distinguish two cases: (a) Fewer then all sites in an $\alpha$-domain are affected by the POVM outcome $K_\alpha$. Then, the domain is simply shrunk, and the graph $G$ is unaffected. (b) All sites in an $\alpha$-domain are affected by the POVM outcome $K_\alpha$. Then, the encoded qubit residing on that domain is measured in the $X$-basis; see Eq.~(\ref{eqn:Pc}).

\begin{table}[tb]
  \begin{tabular}{c|r|r|r}
    %\parbox{1.4cm}
    {POVM outcome\vspace{0mm}} & \multicolumn{1}{c|}{$z$} & \multicolumn{1}{c|}{$x$} & \multicolumn{1}{c}{$y$} \\ \hline
    %\parbox{1.6cm}
    {stabilizer generator} & $\lambda_{i}\lambda_{j} \sigma_z^{[i]}\sigma_z^{[j]}$, & $\lambda_{i}\lambda_{j} \sigma_x^{[i]}\sigma_x^{[j]}$ & $\lambda_{i}\lambda_{j}\sigma_y^{[i]}\sigma_y^{[j]}$ \\
    $\overline{X}$ & $\bigotimes_{j = 1}^{4|{\cal{C}}|} \sigma_x^{[j]}$ & $\bigotimes_{j = 1}^{4|{\cal{C}}|} \sigma_z^{[j]}$ & $\bigotimes_{j = 1}^{4|{\cal{C}}|} \sigma_z^{[j]}$ \\
    $\overline{Z}$ & $\lambda_i \sigma_z^{[i]}$ & $\lambda_i \sigma_x^{[i]}$ & $\lambda_i \sigma_y^{[i]}$
  \end{tabular}
 \caption{\label{tbl:encoding} The dependence of stabilizers and encodings
  on the local POVM outcome. $|{\cal C}|$ denotes the total number of sites contained in a
  domain ${\cal C}$ and $i\&j = 1\, ..\, 4|{\cal{C}}|$  label any two distinct virtual qubits in the same domain ${\cal C}$ (as there are four vitural qubits in a site; see also Fig.~\ref{fig:square}b). The square lattice ${\cal{L}}$ is  bi-partite and all sites can be
  divided into either $A$ or $B$ sublattice, $V({\cal{L}}) = A \cup B$, and $\lambda_i = 1$ if the virtual qubit $i \in v \in A$ and  $\lambda_i = -1$ if
   $i \in v' \in B$. This is due to the negative sign in the stabilizer generator for a singlet $|\phi\rangle_{ij}$, 
   $(-\sigma_{\mu}^{[i]}\sigma_{\mu}^{[j]})|\phi\rangle_{ij}= |\phi\rangle_{ij}$ for an edge $(i,j)$. The logical Y operator can be defined as $\overline{Y}=-i \overline{Z}\overline{X}$.}
\end{table}

\begin{figure}
 \includegraphics[width=0.5\textwidth]{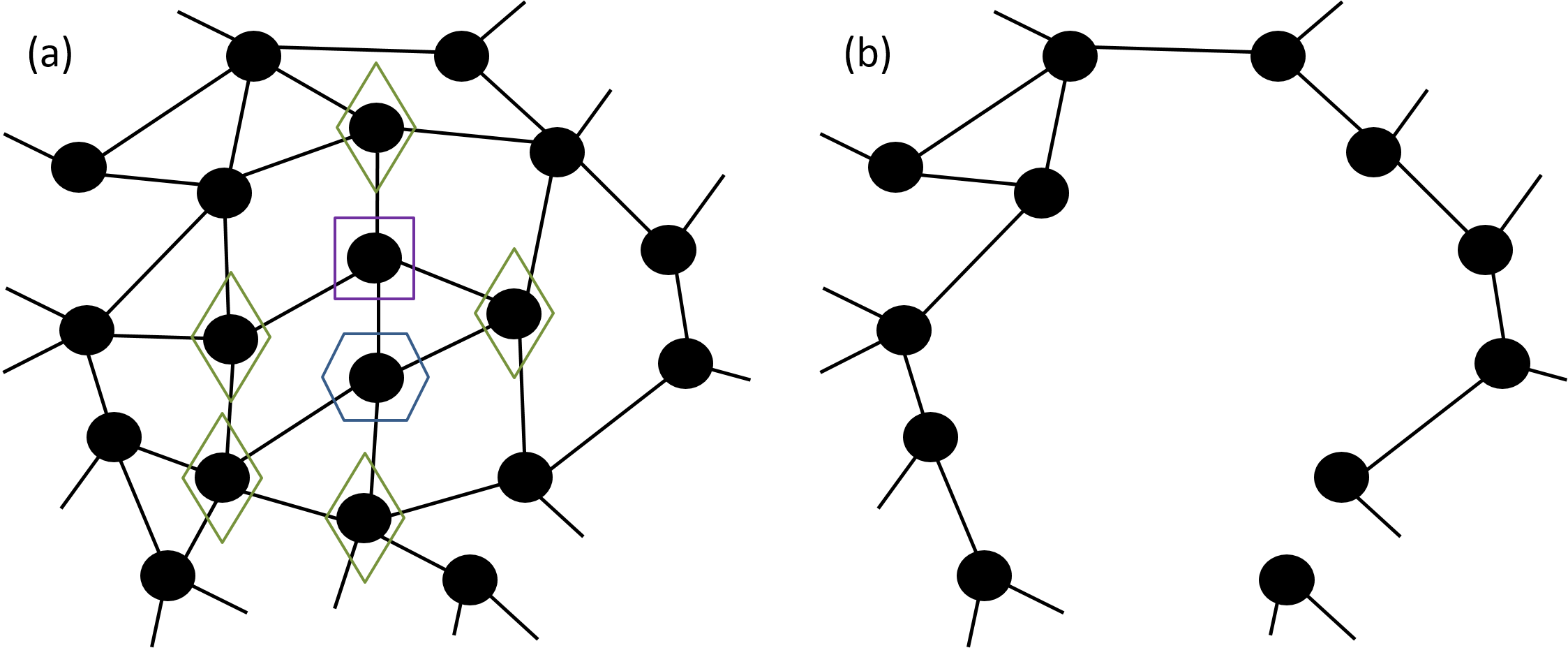}
        \caption{(color online) Part of a random graph for domains (solid circles). (a) The square indicates an $X$-measured domain and the hexagon indicates a $Y$-measured domain. In this example, the two measured domains are neighbors, and the effect on the graph will induce non-planarity. A simple approach is to apply active $Z$ measurement on those domains (indicated by the diamonds) that enclose these connected $X$ or $Y$-measured domains, similar to the game of go.  (b) The upshot of the active $Z$ measurements will remove these $X/Y$-measured domains as well as active $Z$-measured domains but will restore planarity. 
         \label{fig:neighborERR} }
\end{figure}
\begin{figure}
   \includegraphics[width=8cm]{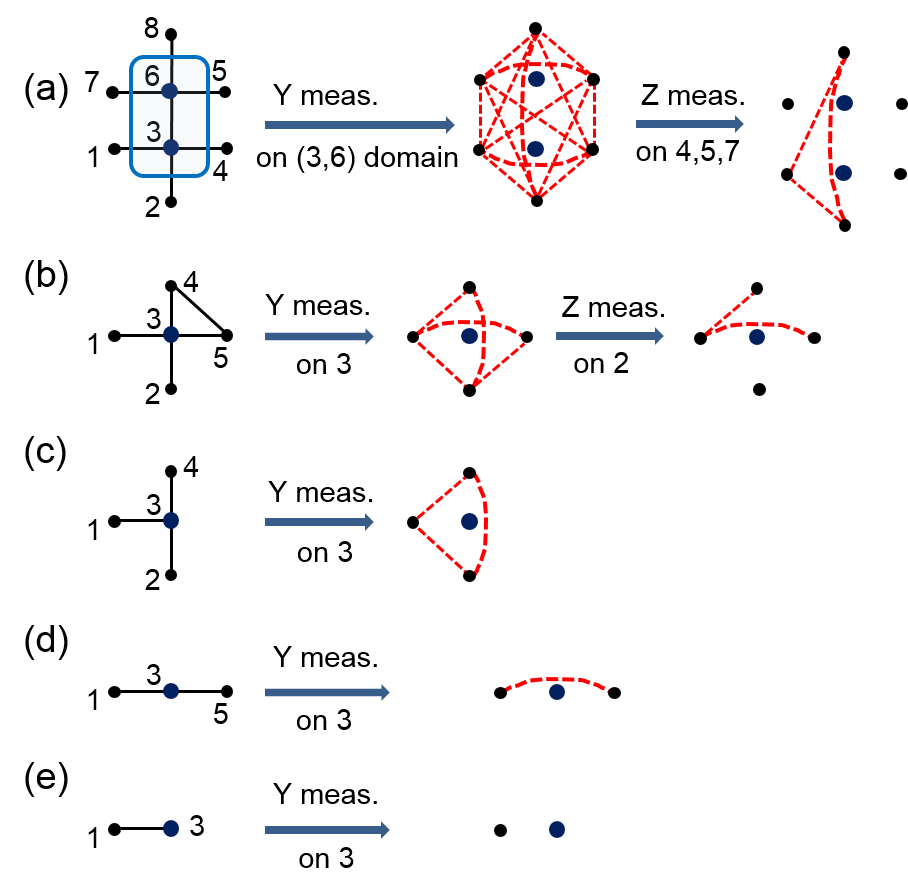}
  \caption{\label{fig:graphrules} (color online) Graph transformation rules on $Y$  measurements on a
  spin-2 domains. (a) illustrates the case where sites 3 and 6 belong to the same domain (indicated by the rectangle) and each of
  them has a $K_\alpha$ outcome such that the effect on the domain is an effective logical $Y$ measurement; others sites are assumed to be of different domain types (i.e. $F_{\beta\ne \alpha}$).
  (b) gives an example where four distinct domains are connected to
  center spin-2 site and  that there is an additional edge between domains 4 and 5.
  To further make the graph remain planar, logical Pauli $Z$ measurements are performed.
  (c) shows the case where three distinct domains are connected to the spin-2 site.
  (This case can arise, e.g., as
  one of the neighboring domains was deleted in  the second step of (a) or (b) associated with
  other spin-2 site.)  (d) and (e) exemplify the cases of, respectively, two and one domain connected
  to a spin-2 site. Note that the above list does not exhaust all possibilities
  (due to other possible connections between neighbors, albeit planar)
   but just serves to illustrate that $Y$ measurement can be
  treated to maintain planarity.
}
\end{figure}

If the latter happens, in terms of CGSB, measurement can be either a logical ${X}$ or a logical $Y$ measurement, and the state resulting from such measurement is again a graph state, and the new graph can be deduced from simple graph rules~\cite{Hein}; see Fig.~\ref{fig:graphrules} for illustration for $Y$-measurement.

\subsection{Restoring planarity by thinning}
\label{sec:thinning}
From the discussions above, we understand that the AKLT state after POVM measurement on all sites is transformed to a graph state. However, the associated graph is generally not planar. We previously established simple criteria for computational universality of random planar graph states \cite{WeiAffleckRaussendorf11,WeiAffleckRaussendorf12}, namely their corresponding graphs need to have a traversing path, and the domains need to be microscopic. The latter requirement for domains to be microscopic was checked numerically in several trivalent lattices~\cite{WeiAffleckRaussendorf11,WeiAffleckRaussendorf12,Wei13} but can be argued to hold using percolation; see Section~\ref{sec:numerics}.

The second step in the computational procedure, after the POVM Eq.~(\ref{POVMspin2}), is therefore a further round of active measurements with the purpose of restoring planarity of the encoded graph state. The non-planarity was caused by the $\ketbra{\phi^-_\alpha}$ outcomes on
all sites in the domains of the graph state $|G_0\rangle$. The thinning procedure degrades the graph state as a potentially universal resource, but it simplifies the universality proof. In other words, these measurements are performed for the sole purpose of simplifying the reasoning.

\begin{figure*}
    \includegraphics[width=6cm]{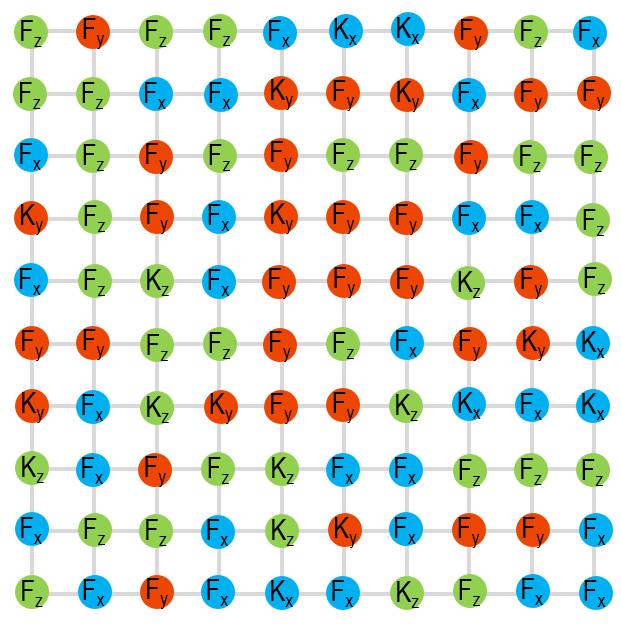}\hspace{1cm}
     \includegraphics[width=6cm]{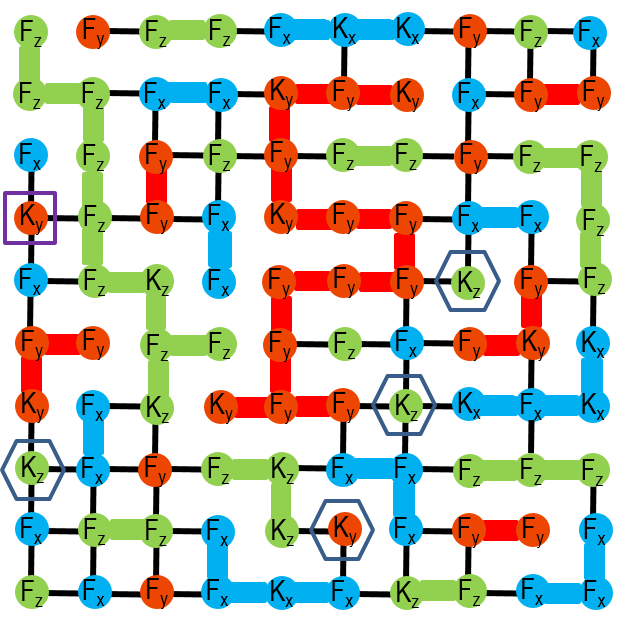}\\ \vspace{0.5cm}
      \includegraphics[width=6cm]{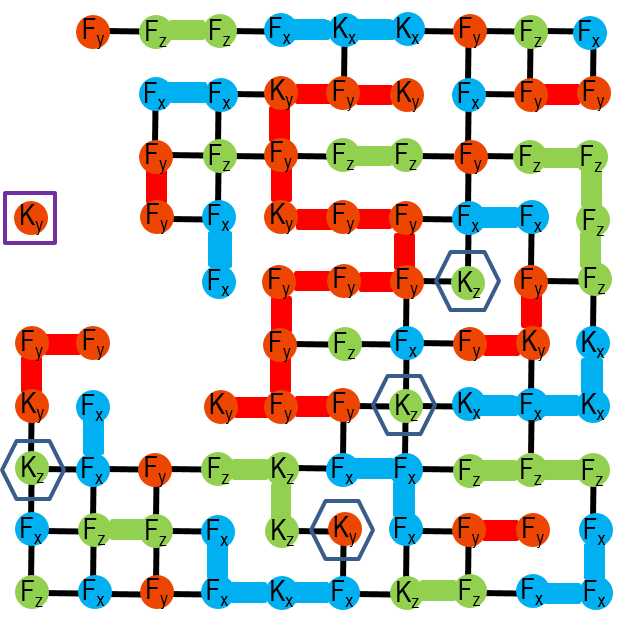}\hspace{1cm}
       \includegraphics[width=6cm]{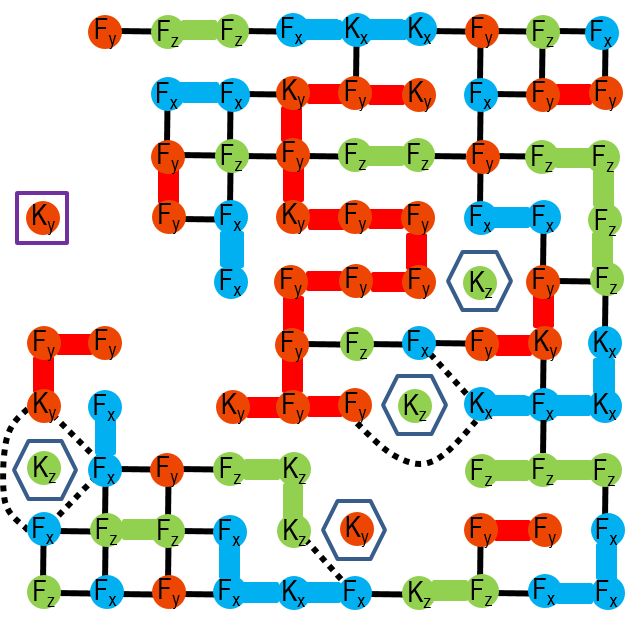}
  \caption{\label{fig:reduction} (color online) Illustration of (1) [top left] $\{F,K\}$ configuration, (2) [top right] domain formation and graph construction (before treating $\{K\}$'s), (3) [bottom left] treating domains which are $X$-measured domains, and (4) [bottom right] treating $Y$-measured domains. Note that the measurement observables are referred to in the CGSB. The result is a plannar graph with fewer domains. In (2) domains are formed by connected sites with same type $\alpha=x,y,z$ (also color-coded) of either $F_\alpha$ or $K_\alpha$. These sites are linked by thicker lines of same color. The edges corresponding to the graph for the graph state of domains. Square boxes indicate the effect of $K$'s on the domain is a logical $X$ measurement (i.e. $X$-measured domain) whereas the hexagon boxes indicate the effect of $K$'s is a logical $Y$ measurement (in the CGSB) or, in other words, Y-measured domains. For those domains that contain at least one $F$, there is no change of the graph due to $K$'s.  Note that for spins on the boundary of the lattice, we imagine, for convenience, that they are attached to spin-1/2 particles (not shown) so as to have four neighbors.
}
\end{figure*}

Our strategy for restoring planarity is to first remove connected POVM ``measured'' (regardless of whether it is $X$- or $Y$-measured) domains  by actively measuring their enclosing/neighboring domains in the logical $Z$ basis, so as to remove these connected ``measured'' domains~\cite{Hein}.
Pauli $Z$-measurements on graph states have the effect of removing the qubit in question from the graph state. On the level of the corresponding graph $G$, the given vertex along with all edges ending in it are removed \cite{Hein}. Hence, $Z$-measurements can be used to excise regions of a graph state. (We remark that the encoded $\overline{Z}$-measurements can be implemented locally on the level of sites of ${\cal{L}}$, as required. This is guaranteed by the coding arising in the given setting, c.f. Table~\ref{tbl:encoding}.) 
Thereby, we remove all $X$-measured domains and  isolated
{\it multi-site\/} (i.e. those with more than 2 sites) $Y$-measured
domains by the same procedure. (We note that the $X$ and $Y$ are referred to in the canonical graph-state basis.) The non-planarity caused by these POVM ``measured'' domains is recovered quasi-locally; see Fig.~\ref{fig:neighborERR}.

\begin{figure}
 \includegraphics[width=0.5\textwidth]{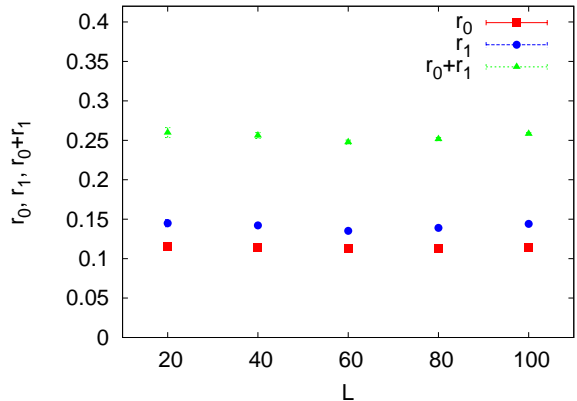}
        \caption{(color online)  The fractions of measured domains  vs. size $L$. The quantity $r_0$  (with ``square'' symbols) is the ratio of the number of both X and Y measured domains to the number of total domains in $|G_0\rangle$, and $r_1$ (with ``circle'' symbols) is the ratio of the number of additionally Z measured domains (due to the thinning procedure) to the number of total domains. ``Triangle'' symbols represent the sum $r_0+r_1$. The fractions are approximately independent of the linear size $L$. 
         \label{fig:RvsL} }
\end{figure}

Then we proceed to deal with the remaining isolated  $Y$-measured
domains which contain either one single or two sites (which can have at most 6 neighboring sites and hence domains). 
The effect of $Y$-measured domains on the graph is to apply
local complementation before removing the vertices corresponding to the $Y$-measured domains. If the $Y$-measured domain has three or fewer neighboring domains, the local complentation still preserves planarity. But when the number of neighbors is four or more, we then actively apply  $Z$ measurement
on some of the neighboring domains (see Fig.~\ref{fig:graphrules}) to maintain
local planarity of the graph. (In principle, any multiple-site
$Y$-type flipped domains can be dealt with. But they appear with a
lower probability.) In Fig.~\ref{fig:reduction}, we give an example from our simulation and show explictly the steps in obtaining a planar graph. In Fig.~\ref{fig:RvsL}, we give the fraction of both the X and Y measured domains (due to the $K$ POVMs) in the graph $G_0$ and the fraction of the additionally Z measured domains needed to restore planarity.

\begin{figure}
 \includegraphics[width=0.5\textwidth]{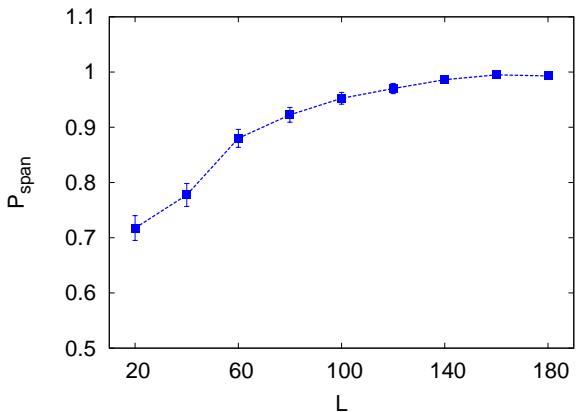}
        \caption{(color online)  The probability of a traversing path, $p_{\rm span}$ vs. the linear size $L$ (with $N=L^2$ the total number of sites). As $L$ increases
        $p_{\rm span}$ also increases. This is obtained with exact sampling. 
         \label{fig:Pspan} }
\end{figure}

In the end we are left with a planar graph
state, whose graph may or may not be percolated. If for large enough
system and with finite nonzero probability, the graphs obtained after the above
procedure are in the supercritical phase, then the resultant graph
states can be used for universal MBQC, implying the original AKLT
state is universal as well. However, if the graphs are not in the
supercritical phase, then it is inconclusive. Our simulations
indicate that we need to use $L$ of order $80$ or larger in order
to show that the graphs are in the supercritical phase with high
probability such as 90\%; see Fig.~\ref{fig:Pspan}. Of course, more sophisticated procedure to deal with
multiple-site $Y$-type domains as well as $X$-type domains can
reduce the size $L$ needed in the simulations, as more vertices will be
preserved. But fortunately, the simple procedure described above turns out to be sufficient.

To carry out the simulations, we still need to sample the configuration $\{F,K\}$ according to the exact distribution $p(\{F,K\})$~\cite{WeiAffleckRaussendorf11}. In Section~\ref{sec:weight} we describe the formula for it and the proof is relegated to Appendix~\ref{sec:proof2}.

\section{Exact weight formula and simulation results}
\label{es}
The exact sampling is needed, as random assignment of $F$ and $K$
POVM outcomes does not correctly reflect the correlation that these
outcomes must obey due to multipartite entanglement in the AKLT
state~\cite{Korepin}. Moreover, many of randomly chosen assignment of $F$ and $K$
are not valid measurement outcomes (as see below by the
incompatibility condition). This latter complication sets the spin-2 case apart from the spin-3/2 case (in addition to the POVM itself). Employing the exact sampling 
also enables us to estimate the probability (at least the lower bound) of
obtaining a universal resource state from performing the reduction
procedure. We note that as long as the reduction procedure gives a
finite, nonzero success probability in the large system limit then the
original state is still regarded as a universal resource state
(though of probabilistic nature). The weight formula that we discuss
below will enable the exact sampling and hence simulations using it
 will give final words on whether the spin-2 AKLT
state on square lattice is a universal resource for MBQC or not.

\subsection{The weight formula}
\label{sec:weight}
\begin{figure}
%\raggedleft
 \includegraphics[width=0.5\textwidth]{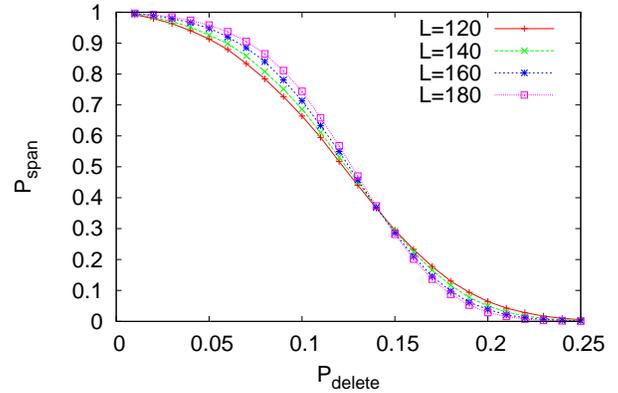}
        \caption{(color online)  The probability of a traversing path $p_{\rm span}$ vs. the probability of deleting each vertex (in the random planar graph after the thinning procedure) $p_{\rm delete}$ 
       for $L=120$ (``circle''), 140 (``square''), 160 (``cross''), 180 (``triangle''), where    $N=L^2$ is the total number of sites. The lines are used to guide the eyes. The threshold of $p_{\rm delete}$ is approximately $0.142(3)$. 
        The crossing for these curves indicates that
        there is a percolation transition from the supercritical to subcritical phase in the thermodynamic limit.
         \label{fig:percoSquare} }
\end{figure}
Let us recapitulate the notations introduced in Sec.~\ref{sec:K}. Consider a spin-2 AKLT state on a bi-colorable lattice ${\cal L}$
(generalization to non-bicolorable lattices is possible), and POVM
elements $F_\alpha$ and $K_\alpha$ ($\alpha=x,y,z$). Denote by $J_F
\subset {\cal L}$ the set of sites where the POVM outcome is of
$F$-type and by $J_K={\cal L}\backslash J_F$ the set of sites where POVM
outcome is of $K$-type. Here additionally we denote by $D_K$ the set of {\it domains} where
the number of $K$-type POVM elements is equal to the total number of sites in the domain. 
Denote $\{F,K\}$ the set of POVM outcomes corresponding to $F_{\alpha(v)}
^{(v)}$ and $K_{\beta(w)}^{(w)}$ and the probability for such occurrence
is $p(\{F,K\})$.

We have also introduced the graph state $|G_0\rangle$ in Eq.~(\ref{eqn:G0}), and we will denote its stablizer group as ${\cal S}(|G_0\rangle)$. 
The stablizer operators ${\cal K}_\mu$'s for $|G_0\rangle$ with respect to the encoding in Table~\ref{tbl:encoding} can be either $\pm X_c \bigotimes_{\mu \in {\rm Nb}(c)} Z_\mu$ or $\pm Y_c \bigotimes_{\mu \in {\rm Nb}(c)} Z_\mu$ (see Appendix~\ref{sec:VAstablizer}), where ${\rm Nb}(c)$ denotes the set of neighbors of vertex $c$. 
 As explained in Sec.~\ref{sec:K}, the effect  of $K$-type POVM elements on a strict subset of sites in a domain only shrinks the size of a domain, whereas $K$-type POVM measurement on {\it all\/} sites
in a domain in $D_K$ amounts to the measurement (on $|G_0\rangle$)
of an encoded logical ${X}$.  In the CGSB, all stabilizer operators are of the form  $X_c \bigotimes_{\mu \in {\rm Nb}(c)} Z_\mu$, but then the effect of $K$-type POVM elements 
in a domain inside $D_K$ amounts to the measurement
of an encoded observable either ${X}$ or ${Y}$, depending on the existence of a self-loop. Let us label the set of all domains (i.e. vertices of the graph $G_0$) by $V$, the set of all inter-domain edges in ${\cal L}$ by ${\cal E}$ and the set of all edges of $G_0$ by $E$. Note that $E$ is obtained from ${\cal E}$ by a modulo-2 operation~\cite{WeiAffleckRaussendorf11}.
Now we introduce a $|V|\times |D_K|$ binary-valued matrix $H$  with its entries defined as follows,
\begin{subequations}
\begin{eqnarray}
&&H_{\mu\nu}=0, \ \mbox{if} \, [{\cal K}_\mu,O_\nu]=0,\\ 
&&H_{\mu\nu}=1, \ \mbox{if} \, \{{\cal K}_\mu,O_\nu\}=0,
\end{eqnarray}
\end{subequations}
where ${\cal K}_\mu$ is the stabilizer operator associated with the vertex (or domain) $\mu\in V$ of the graph $G_0$, $O_\nu$ is the operator defined in Eq.~(\ref{eqn:Pc}), and $\nu\in D_K$; see also Appendix~\ref{sec:VAstablizer}. Let $\dK$ denote the dimension of the kernel of matrix $H$.
The utility of $H$ is in the following lemma.

\begin{Lemma} If there exists a set $Q \subset D_K$ such that $-\mathop{\otimes}_{\mu\in Q} O_\mu \in {\cal S}(|G_0\rangle)$, then $p(\{F,K\})=0$. Otherwise,  
\begin{equation}
\label{eqn:pWeight}
p(\{F,K\})=c \,\left(\frac{1}{2}\right)^{|{\cal E}|-|V|+2 |J_K|-\dK},
\end{equation}
where $c$ is a constant.
\end{Lemma}
We subsequently refer to the above condition for $p(\{F,K\})=0$ as the {\it incompatability condition\/}.
The incompatibility condition implies that not all POVM outcomes labeled by $F_\alpha$ and $K_\alpha$ can occur. When there is no $K$ outcome, Eq.~(\ref{eqn:pWeight}) reduces to $p=c\, 2^{|V|-|{\cal E}|}$ of previous results~\cite{WeiAffleckRaussendorf11}. The correlation of $F$'s and $K$'s at different sites is reflected either in the incompatibility condition (if it is met) or else in the factor $\dK$.  The probability distribution of $\{F,K\}$ is thus very far from being independent and random. For the proof of the lemma, see Appendix~\ref{sec:proof2}.

With the weight formula we can sample the exact distribution of physically allowed POVM outcomes $\{F,K\}$ and carry out the procedure to restore planarity of the random graphs associated with the post-POVM states. To show that these graph states are universal, we need to show that (i) the domains are not of macroscopic size and (ii) these graphs reside in the supercritical phase (as the system size increases). For (i), we remark that the largest domain size can only be logarithmic with
the system size $N$. This was confirmed earlier in our simulations
of the honeycomb case. In fact, it is well-known from percolation
that below the percolation threshold, the largest connected
structure is logarithmic in $N$ (by a previous study of percolation by Bazant~\cite{Bazant}). This can be applied to our
present study.  If we only sample $F$'s, there are three types of
outcome $x, y, z$ and
 locally with the probability 1/3. Domains consist of connected sites of same type. For each type,
 it is a site percolation problem, at the occupation probability 1/3, smaller than the site
  percolation threshold 0.59 (cf. for honeycomb, the threshold is 0.69). This means that each type of domain can be at
   largest logarithmic in $N$. Because $1/3$ is very far from $0.59$ and the AKLT state has a small and finite correlation length,
   inclusion of correlation will not change the scaling.  Furthermore,
   the $K$'s can only decrease the domain size, so the relevant largest domain size is for sampling $F$'s only.

The sampling is obtained by using the standard Metropolis algorithm for updating $\{F,K\}$ configurations. One notable distinction is that we need to check whether the next configuration satisfies the incompatibility condition. If it is satisfied,  we  then abondon that configuration and generate another one until the incompability condition is not satisfied. Then the configuration is accepted using the standard probability ratio, as was done in e.g. the spin-3/2 AKLT case~\cite{WeiAffleckRaussendorf11}.

\subsection{Numerical simulations}
\label{sec:numerics}
\begin{figure}
 \includegraphics[width=0.5\textwidth]{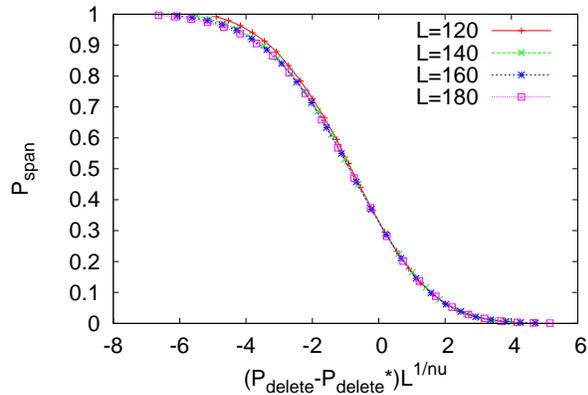}
        \caption{(color online)  Data collapse in $p_{\rm span}$ vs. $(p_{\rm delete}-p_{\rm delete}^*)\cdot L^{1/\nu}$ for different $L=120$ (``circle''), 140 (``square''), 160 (``cross''), 180 (``triangle''), where $N=L^2$ the total number of sites. The lines are used to guide the eyes. We note that
        $p_{\rm span}$ represents the probability of a traversing path and $p_{\rm delete}$  the probability of deleting each vertex in the random planar graph after the thinning procedure.  The $p_{\rm delete}^*\approx 0.142$ is extracted from the results in Fig.~\ref{fig:percoSquare}, and $\nu=4/3$  is used from 2D percolation. There is no tuning parameter. Data collapse is clearly seen (except for small deviation in the data from $L=120$, possibly due to the small size) and this confirms  that there is a continuous phase transition.
         \label{fig:collapse} }
\end{figure}
   
%    and $p_{\rm delete}$  the probability of deleting each vertex in the random planar graph after the thinning procedure. 
%   Regarding point (ii),  we compute the probability of the spanning cluster, $p_{\rm span}$, at $p_{\rm
%   delete}=0$ (i.e. without removing any vertex from the random graphs) for different $L$, and

  Here, we first take typical random planar graphs after the thinning procedure and check the probability  
        $p_{\rm span}$ that a traversing path exists.  We find that $p_{\rm span}$ increases as the linear size $L$ of the square lattice increases and approaches to unity eventually; see Fig.~\ref{fig:Pspan}. This suggests that for $L$ large enough, the random graphs resulting from the thinning procedure are percolated.  To confirm this and examine the connectivity of the typical random graphs and
         perform site-percolation simulations, by removing every vertex in the graphs with a probability $p_{\rm delete}$.  It is important to note that we are interested in the connectivity for the random planar graphs resulting from the thinning procedure, i.e. those graphs at $p_{\rm delete}=0$, but as a means to characterize the robustness in the connectivity we examine how much we have to delete the vertices in order to remove all the traversing paths that were there and then use the obtained threshold value $p_{\rm delete}^*$ as the quantification. We expect that when $p_{\rm delete}$ is sufficiently large, there will not be any traversing path left. Indeed this is what we observe. Additionally, the crossing of curves in Fig.~\ref{fig:percoSquare} for different sizes indicates that there is a percolation transition (at $p_{\rm delete}^*\approx 0.142$) from the supercritical to subcritical phase in the thermodynamic limit. This then justifies the separation 
of two phases: (a) supercritical phase of percolation when $p_{\rm delete} < p_{\rm delete}^*$, where there are macroscopic number of traversing paths; and (b) subcritical phase of percolation when $p_{\rm delete} > p_{\rm delete}^*$, where no traversing path exists. Graphs residing in the supercritical phase of percolation contain a macoscopic number of traversing paths. This shows that our system sitting at $p_{\rm delete}=0$ can be used to generate a network of entanglement that is universal for measurement-based quantum computation.

As a further confirmation of the transition, we can rescale the horizontal axis via $(p_{\rm delete}-p_{\rm delete}^*) L^{1/\nu}$ and collapse all data points approximately on a universal curve; see Fig.~\ref{fig:collapse}. The exponent $\nu=4/3$ is used for the correlation-length critical exponent of the 2D percolation universality class. There is no tuning parameter used and the data collapse demonstrates that the phase transition is continuous. By establishing the transition and the critical point, we have thus showed that the random graph states we generated possess graphs residing in the supercritical phase of percolation and hence these random graph states are universal for MBQC~\cite{WeiAffleckRaussendorf12}. Therefore, the spin-2 AKLT state, from which the random graph states are generated by local measurement, is itself a universal resource.
\subsection{AKLT state on the diamond lattice}
The technique for spin-2 square lattice can be applied to other
planar lattices with the coordination number  $z=4$ and be extended to 3D ones, such as the diamond  lattice.  For the latter lattice one can
generalize the consideration to three-dimensional percolation and consider conversion of a
3D random graph state (in the supercritical phase) to a 3D cluster state, which can then
be useful for providing a fault-tolerant implementation of MBQC~\cite{TO}.  Even though we do not pursue technical demonstration in the present paper,  we believe that the AKLT state on the diamond lattice is universal for MBQC. 

First of all, as the site percolation threshold of the diamond lattice is approximately $0.430$, larger than $1/3$, the largest domain size will be at most logarithmic in the total number of the spins, and hence is not macroscopic. Secondly, the planarity that was the key obstacle in 2D is now relaxed to three-dimensionality. Thus fewer domains need to be actively measured in the canonical $Z$ basis and the connectivity can be more easily preserved. The resulting 3D random graph states are thus expected to possess sufficient connectivity in terms of percolation. From there, the reduction of 3D random graphs to 3D regular graphs could be carried out by local measurement. The sufficient connectivity implies that measurement-based quantum computation can be realized. The main advantage of using such a 3D resource state is that the quantum computation can be implemented in a topologically protected manner and can tolerate high error rates~\cite{TO}. We remark that there exist $z=3$   regular structures in 3D with higher percolation thresholds than $1/3$~\cite{Tran}.  AKLT states on these lattices also support fault-tolerant quantum computation, but these 3D lattices with $z=3$ are not as natural as the diamond lattice.

\section{Deformed AKLT and transition in quantum computational power}
\label{sec:Deformed}
In this section we discuss how our techniques and results can be extended to inquire the question whether there is an extend region in the Hilbert space surrounding the AKLT point such that the states can also provide universal resource. Is there a transition in terms of quantum computational power? In the following we shall provide argument to support the affimative answers without carrying out detailed numerical simulations. 

Let us
define a deformation operator
\begin{eqnarray}
\label{eqn:DeformedWF}
&& {\cal D}(a)\equiv \frac{1}{a}\big(\ketbra{S_z\!=\!2}+\ketbra{S_z\!=\!-2}\big)+ \\
               && \ketbra{S_z\!=\!1}+\ketbra{S_z\!=\!-1}+\ketbra{S_z\!=\!0},\nonumber
\end{eqnarray}
where we shall restrict to the case $a\ge 1/\sqrt{3}$. The operator ${\cal D}(a)$ is then used to deform the original spin-2 AKLT Hamiltonian into
\begin{equation}
\label{eqn:DeformedH}
H_{\rm D}(a)=\sum_{\langle i,j\rangle} [{\cal D}(a)_i\otimes{\cal D}(a)_j]h_{i,j}^{\rm AKLT} [{\cal D}(a)_i\otimes {\cal D}(a)_j],
\end{equation}
where $h_{i,j}^{\rm AKLT}$'s are the two-body terms between pairs of neighboring sites $\langle i,j\rangle$ in the original AKLT Hamiltonian. Similar deformation was discussed in the spin-1 and spin-3/2 cases~\cite{DarmawanBrennenBartlett,NiggemannKlumperZittarz,Verstraete04}. It is easy to see that the following one-parameter deformed AKLT state
\begin{equation}
|\psi_D(a)\rangle= [{\cal D}^{-1}(a)]^{\otimes N} |\psi_{\rm
AKLT}\rangle
\end{equation}
is the ground state of the deformed Hamiltonian~(\ref{eqn:DeformedH}), and it reduces to the original AKLT state at $a=1$. Moreover, when $a\gg 1$, $|\psi_D(a)\rangle$ is essentially the N\'eel ordered state, or more precisely, a superposition of two different N\'eel patterns, $|S_z=3/2\rangle\otimes |S_z=-3/2\rangle \otimes \cdots$ and $|S_z=-3/2\rangle\otimes |S_z=3/2\rangle \otimes \cdots$, which does not have sufficient entanglement for MBQC, similar to the spin-3/2 case~\cite{DarmawanBrennenBartlett}. As the spin-2 AKLT state on the square lattice (and the diamond lattice) is not N\'eel ordered, the wavefunction in (\ref{eqn:DeformedWF}) must possess a transition point in terms of phases of matter at certain value of $a$.

The questions that concerns us here is whether there is a finite region around the AKLT point $a=1$ such that the wavefunction~(\ref{eqn:DeformedWF}) is still universal for MBQC. (Another interesting question is whether the disappearance of the universality coincides with the valence-bond-solid to N\'eel transition, but we leave this for future investigation.) We shall argue below that indeed there is a finite region around the AKLT point such that the ground state is universal.

First of all, due to the deformation, the POVM in~(\ref{POVMspin2}) cannot be directly applied but can be modified to work,
\begin{subequations}
\label{deformedPOVMspin2}
\begin{eqnarray}
\tilde{F}_x(a)&=&F_x {\cal D}(a), \ \tilde{F}_y(a)=F_y {\cal D}(a), \\
\tilde{F}_z(a)  &=&\sqrt{\frac{3-a^2}{2a^2}}\,F_z {\cal D}(a),\\
\tilde{K}_\alpha(a)& =&K_\alpha {\cal D}(a), 
\end{eqnarray}
\end{subequations} 
with which one can verify that $\sum_\alpha \tilde{F}^\dagger_\alpha(a) \tilde{F}_\alpha(a) +\sum_\alpha
\tilde{K}^\dagger_\alpha(a) \tilde{K}_\alpha(a)=\openone$.  The physical intuition of how the above POVM can work for the purpose of MBQC is that it can be regarded as a two-step process: (i) first, the deformation ${\cal D}(a)$ undoes the action ${\cal D}^{-1}(a)$ in $|\psi_D(a)\rangle$ and converts it back to the AKLT state; (ii) the original POVM $\{F_\alpha,K_\beta\}$ then acts on the AKLT state as before.

Furthermore, the weight formula in Lemma 1 is then modified to
\begin{equation}
p(\{\tilde{F}_\alpha,\tilde{K}_\beta\})= c(a) \left(\frac{3-a^2}{2a^2}\right)^{n_{F_z}} p(\{F,K\}),
\end{equation}
where $c(a)$ is an $a$-dependent normalization that is independent of $\{\tilde{F},\tilde{K}\}$, $n_{F_z}$ denotes the total number of sites where the POVM outcome is $F_z$, and $p(\{F,K\})$ is given by Eq.~(\ref{eqn:pWeight}) and the incompatibility condition remains the same.
The only difference in the weight formula is the additional factor involving $n_{F_z}$ and when $a=1$ the formula reduces to Eq.~(\ref{eqn:pWeight}). As we have found a finite percolation threshold $p_{\rm delete}^*$ for $a=1$ case and the weight formula is continuous in $a$, we thus expect to have $a$-dependent threshold $p_{\rm delete}^*(a)$, except when the largest domain size becomes macroscopic. Thus there must exist a finite range around $a=1$ such that the $p_{\rm delete}^*$ is nonzero and the largest domain size is microscopic. Therefore, the deformed AKLT state~(\ref{eqn:DeformedWF}) is universal for MBQC in this region, and there exists a transition of quantum computational power as $a$ increases, from being universal to non-universal. This argument also applies to the three dimensional case, and in particular, the diamond lattice.

Even though we do not carry out simulations here to pin down the exact transition, we describe what may be achieved with present techniques. By studying the dependence of the largest domain size on the parameter $a$, we can extract at least the upper bound on the transition, $a_{\rm up}$, when the largest domain size begins to become macroscopic. Via the dependence of $p_{\rm delete}^*(a)$ and the location, $a_{\rm lower}$, at which $p_{\rm delete}^*(a_{\rm lower})$ becomes zero, we can extract the lower bound of the transition. The exact transition $a_{\rm trans}$ of the quantum computational power will then lie between the two bounds: $a_{\rm lower}\le a_{\rm trans}\le a_{\rm up}$. One could also determine the spontaneous staggered magnetization to extract the transistion of the phase of the matter. It would be interesting to see how the two transitions differ.

\section{Concluding remarks}
\label{Sum}

The family of Affleck-Kennedy-Lieb-Tasaki states provides a
versatile playground for universal quantum computation. The merit of
these states is that by appropriately choosing boundary conditions
they are unique ground states of two-body interacting Hamiltonians,
possibly with a spectral gap above the ground states. Here we have overcome several obstacles and shown that the spin-2 AKLT state on the square lattice is also a universal resource for measurement-based quantum computation. In particular, we were able to derive an exact weight formula for any given POVM outcome. Combined with a thinning procedure to restore planarity of random graph states, we performed Monte Carlo simulations and demonstrated that the assoicated planar random graphs from the procedure possess sufficient connectivity and reside in the supercritical phase. In particular, our numerical site-percolation simulations showed that as the deletion probability increases (i.e., as the occupation probability decreases) the system of the above random graphs makes a continuous phase transition from the supercritical phase of percolation to the subcritical phase of percolation. Moreover, the continuous phase transition is consistent with the universality class of the 2D percolation. These demonstrate that typical random graph states obtained via our local measurement procedure on the AKLT state are universal for measurement-based quantum computation. Thus, the spin-2 AKLT state on its own is also universal.  

One of the important enabling points for our proof is the spin-2 POVM. This POVM was used previously by us in considering hybrid AKLT states with isolated spin-2 and lower-spin entities~\cite{WeiEtAl}.  Here, we were able to deal with the case that spin-2 particles are neighbors.   
Another enabling point is the exact weight formula that we derived for arbitrary measurement outcomes according to a spin-2 POVM on all spins.
This formula can be extended to Pauli measurements on stablizer states~\cite{Elliott}, which may be useful for classical simulations of certain gates. Our weight formula also demonstrates the most pronounced difference between 
the spin-3/2 and spin-2 cases: for spin-3/2 AKLT on bi-partite lattices, all possible combinations of POVM outcomes do indeed occur with non-zero probability, whereas,  for the spin-2 case, certain combinations of POVM outcomes do not occur, i.e., have probability zero. This shows that the POVM outcome in spin-2 case is very different from being random and independent, and the simulations from the exact weight formula are crucial in answering whether the spin-2 AKLT state is universal for MBQC or not.

The emerging picture from
our series of study on the quantum computational universality in the two-dimensional
AKLT valence-bond family
 is as follows.  AKLT states involving
 spin-2 and other lower spin entities are universal if they reside on
 a two-dimensional frustration-free regular lattice with any
 combination of spin-2, spin-3/2, spin-1 and spin-1/2 (consistent
 with the lattice). Additionally, the effect of frustrated lattice may not be serious and can always be
 decorated (by adding additional spins) such that the resultant AKLT
 state is universal. We conjecture that the result hold in three dimensions as well.

We remark that an alternative approach to demonstrate the quantum computational universality is to explicitly construct universal gates and show that they can be realized to simulate arbitrary quantum circuits. This can be done via the so-called correlation-space approach of MBQC~\cite{Gross}, as demonstrated in the spin-3/2 case~\cite{Miyake11}. One could carry out this procedure for the spin-2 case. However, similar to the spin-3/2 case, the key enabling ingredient is the POVM~(\ref{POVMspin2}), as well as the thinning procedure in Sec.~\ref{sec:thinning}. After these steps, the state becomes a graph state, and thus the gate construction in the correlation space can be implemented accordingly. Moreover, the question of whether there are sufficient gates that can be used to simulate arbitrary quantum circuits may still rely on the percolation argument and the numerical simulations done here.  However, it may be possible to bypass the thinning procedure and directly design gates to ensure that arbitrary quantum circuit can be simulated. But here we do not pursue this alternative approach.

Simple states that are ferromagnetic or  N\'eel ordered are believed not to possess entanglement structure useful for MBQC. AKLT states in 2D were shown to be disordered, i.e., not N\'eel ordered, and some (not including the diamond lattice) in 3D were N\'eel ordered~\cite{Param}.  Could it be that being disordered (and non-frustrated) is a sufficient condition for these AKLT states being universal?  We believe it is not. In terms of spin magnitude, one expects that the larger the spin magnitude the more classical the system becomes, i.e., the effect of non-commutativity of spin operators becomes less and less important. For larger spin magnitdue $S$, despite being disordered, the system is becoming more classical. On this ground we suspect that for large enough $S$, AKLT states will eventually cease to be  universal for MBQC. But what is the boundary of being universal and non-universal and what would be the decisive physical properties for determining such a boundary?  To address this question requires further insight. Nevertheless, our results in the present paper show that the $S=2$ case is still in the universal regime.

One direction of generalization is to investigate the robustness of the resource under small perturbations, e.g., slightly away from the AKLT Hamiltonian. We have discussed a particular deformation of the spin-2 AKLT Hamiltonian and  the transition in quantum computational power in Sec.~\ref{sec:Deformed}. One could also consider deformations that take the spin-2 kagom\'e AKLT state to an effective cluster state, turning a non-universal state to a universal one. It would also be interesting to consider the effect of finite temperature and at how high the temperature the system considered here can still support a universal resource~\cite{WeiLiKwek}.  This is beyond the scope of the present paper and is left  for future investigation.  Another direction one can study is the connection of the resourcefulness of the AKLT states (and their generalization) to the symmetry-protected topological (SPT) order, the connection of which was recently found in one dimension~\cite{ElseSchwarzBartlettDoherty}. AKLT states serve as concrete examples of SPT ordered states. However, what symmetry is needed and in what SPT phases can there be naturally protected universal gates? Progress along this direction in two and higher dimensions can potentially advance our understanding towards complete characterization of universal resource states.  We also leave this intriguing question for future investigation.

We wish to conclude by commenting on potential experimental implementations. Bosons with multiple hyperfine states can be used for high spins, and thus they are a natural candidate to realize possible AKLT-like Hamiltonians when placed in optical lattices~\cite{Stamper-KurnUeda,LianZhang}. The difficulties lie in (1) tuning interactions to the desired Hamiltonian regime and (2) local controllability of single atoms. Significant experimental progress has been made in (2) that it is now possible to detect and image single atoms~\cite{Greiner,Sherson}.  One advantage of utilizing cold atoms in optical lattices is the ability to scale up the system. In addition to such a top-down approach, it is also possible to build the resource states from bottom up, such as with entangled photons. The first experimental demonstration of implementing gates in the 1D AKLT state was carried out in the photonic system~\cite{Resch}. The key idea relies on the mapping of the general AKLT state to a bosonic  state (which has a Laughlin-like wavefunction in the coherent-state representation)~\cite{ArovasAuerbachHaldane}, in that
\begin{equation}
|\psi_{\rm AKLT}({\cal L},M)\rangle\sim \prod_{\langle i,j\rangle} (a_i^\dagger b_j^\dagger - b_i^\dagger a_j^\dagger)^M |0\rangle, 
\end{equation} 
where $|0\rangle$ is the vacuum state,  $a_i^\dagger$ creates a boson of type $a$ (such as the horizontal polarization $H$ of a photon) at site $i$, $b_j^\dagger$ creates a boson of type $b$ at site $j$ (such as the vertically $V$ polarizaed photon), and $M$ denotes the number of singlets, with $M=1$ being the original AKLT state. Therefore, creating an AKLT state is equivalent to placing singlet pairs of photons, such as $|HV\rangle-|VH\rangle$, according to the valence-bond construction~\cite{AKLT} and the symmetrization of the photons are automatic~\cite{Darmawan}. In the 1D AKLT state, there are two photons ``per site'' (or per mode), and the measurement of the effective spin-1 degree of freedom can be carried out if the two photons are indistinguishable~\cite{Steinberg07,Steinberg08}. There are three and four photons per site on the 2D honeycomb and square lattices, respectively.  Measurement on multiple indistinguishable photons that is equivalent to measurement in the spin basis can be carried out similarly~\cite{Steinberg07,Steinberg08}.   It is thus necessary to make those photons on the same site indistinguishable, i.e., to achieve good mode matching. A proposal was made in realzing the spin-3/2 AKLT state on the honeycomb lattice and in implementing the key POVM~\cite{LiuLiGu}.  Its generalization to the spin-2 case on the squre and diamond lattices is possible. The advantage of using entangled photons is that small-scale implementation is already within the reach of current technology, but the disadvantage would be to scale up to a large system. Beyond the atomic-molecular-optical schemes, it is very recently proposed  to realize AKLT and general valence-bond states in solid-state systems with, e.g., $t_{2g}$ electrons in Mott insulators~\cite{Sela}. 

\medskip \noindent {\bf Acknowledgment.}  T.-C.W. thanks Chris Herzog for providing access to his workstation, on which a major part of the simulations reported here were carried out. This work was supported by the
National Science Foundation under Grants No. PHY 1314748 and No. PHY
1333903 (T.-C.W.) and by NSERC, Cifar and IARPA (R.R.).

\appendix
\section{The POVM expressed in terms of four-qubit representation}
\label{GR}

Expressed in terms of the four virtual qubits representing a spin-2
particle, the elements  that comprise the spin-2 POVM are
\begin{subequations}
\label{POVMspin2b}
  \begin{eqnarray}
\!\!\!\!\!\!\!\!\!\!{F}_{z}&=&\sqrt{\frac{2}{3}}(\ketbra{0^{\otimes 4}}+\ketbra{1^{\otimes 4}}) \\
\!\!\!\!\!\!\!\!\!\!{F}_{x}&=&\sqrt{\frac{2}{3}}(\ketbra{+^{\otimes 4}}+\ketbra{-^{\otimes 4}})\\
\!\!\!\!\!\!\!\!\!\!{F}_{y}&=&\sqrt{\frac{2}{3}}(\ketbra{i^{\otimes 4}}+\ketbra{(-\!i)^{\otimes 4}})\\
\!\!\!\!\!\!\!\!\!\!{K}_{z}&=&\sqrt{\frac{1}{3}}\ketbra{{\rm GHZ}^-_z} \\
\!\!\!\!\!\!\!\!\!\!{K}_{x}&=&\sqrt{\frac{1}{3}}\ketbra{{\rm GHZ}^-_x}\\
\!\!\!\!\!\!\!\!\!\!{K}_{y}&=&\sqrt{\frac{1}{3}}\ketbra{{\rm
GHZ}^-_y},
\end{eqnarray}
\end{subequations}
where $|\psi^{\otimes 4}\rangle$ is a short-hand notation for
$|\psi,\psi,\psi,\psi\rangle$, equivalent to an eigenstate $|S=2,
S_\alpha\rangle$ of the spin-2  operator $\hat{S}_\alpha$ with an
eigenvalue $S_\alpha=\pm 2$ in either $\alpha=$ x, y, or z
direction. Note that strictly speaking we should use different notations for these POVM elements, but since they are equivalent to those in Eq.~(\ref{POVMspin2}), we retain the same notations. Moreover, $\sigma_z|0/1\rangle=\pm |0/1\rangle$, $\sigma_x|\pm\rangle=\pm|\pm\rangle$, and $\sigma_y|\pm i\rangle=\pm|\pm i\rangle$. The first three elements in Eq.~(\ref{POVMspin2b}) are similar to those in spin-3/2
sites, except the number of virtual qubits being four, and
correspond to good outcomes of type x, y and z, respectively.
Associated with the last three elements, $\ket{{\rm GHZ}^-_z}\equiv
(\ket{0000}-\ket{1111})/\sqrt{2}$, $\ket{{\rm
GHZ}^-_x}\equiv(\ket{++++}-\ket{----})/\sqrt{2}$, and $\ket{{\rm
GHZ}^-_y}\equiv(\ket{i,i,i,i}-\ket{-\!i,-\!i,-\!i,-\!i})/\sqrt{2}$
are the corresponding states and they will be regarded as unwanted
outcomes of type x, y, and z, respectively.  But these GHZ states
are at least eigenstates for certain product combination of Pauli
operators. It can be verified that $\sum_\alpha
{F}^\dagger_\alpha {F}_\alpha +\sum_\alpha
{K}^\dagger_\alpha  {K}_\alpha=P_S$, where $P_S$ is
the projection onto the symmetric subspace of four qubits, i.e.,
identity in the spin-2 Hilbert space. Let us also note that the
${K}$'s operators can be rewritten as
\begin{equation}
\label{eqn:K0} {K}_\alpha= \sqrt{1/2} \ketbra{{\rm
GHZ}_\alpha^-}{F}_\alpha. \end{equation}
Furthermore,  there is a useful relation:  
\begin{equation}
 \ketbra{{\rm
GHZ}_\alpha^-}= \Pi_\alpha\frac{(1-\sigma_{b_\alpha}^{[v;1]}\sigma_{b_\alpha}^{[v;2]}\sigma_{b_\alpha}^{[v;3]}\sigma_{b_\alpha}^{[v;4]})}{2}\Pi_\alpha,
 \end{equation}
where $\Pi_\alpha$ is a projection to a two-dimensional subspace and is an identity operator on the code subspace (for the corresponding POVM outcome); specificially, $\Pi_x=\ketbra{++++}+\ketbra{----}$, $\Pi_y=\ketbra{i,i,i,i}+\ketbra{-i,-i,-i,-i}$ and $\Pi_z=\ketbra{0000}+\ketbra{1111}$.  The label $b_\alpha$ denotes the corresponding type $b$ if $a_c=\alpha$; see Table~\ref{tbl:b}.

\section{The exact form of stabilizer generators}
\label{sec:VAstablizer}

In this section we give the explict form of the stabilizer operator ${\cal K}_{C}$ for the domain labeled by ${C}$. It includes all subtle plus and minus signs. The result is general for all states $|G_0\rangle\sim\bigotimes_{v\in {\cal L}} {F}_{\alpha_v,v}|\psi_{\rm
AKLT}\rangle$, where $F$'s can be of arbitrary spins. This was already considered in the case of the spin-3/2 AKLT state~\cite{WeiAffleckRaussendorf11}, but the argument used there applies more generally.
 
 Let us first explain the notation. Consider a central vertex ${{C}} \in V(G_0(\{F\}))$ and all its
neighboring vertices ${{C}}_\mu \in V(G_0)$. Denote the POVM
outcome for all ${\cal{L}}$-sites $v \in {{C}}, {{C}}_\mu$ by $a_c$
and $a_\mu$, respectively. Denote by $E_\mu$ the set of ${\cal{L}}$-edges that
run between ${{C}}$ and ${{C}}_\mu$. Denote by $E_c$ the set of
${\cal{L}}$-edges internal to ${{C}}$. Denote by $V_c$ the set of all
qubits in ${{C}}$, and by $V_\mu$ the set of all qubits in
${{C}}_\mu$. (Recall that there are 4 qubit locations per
${\cal{L}}$-vertex $v \in {{C}},{{C}}_\mu$.)
 Extending Eq.~(33) of Ref.~\cite{WeiAffleckRaussendorf12} to the spin-2 case, we have
\begin{eqnarray}
\!\!\!\!\!\! {\cal K}_{{C}}& =&
\bigotimes_{\mu}\bigotimes_{e \in E_\mu}
  (\!-\!1)\sigma_{a_\mu}^{(u(e))} \sigma_{a_\mu}^{(v(\!e\!))}
   \bigotimes_{e' \in E_c} (\!-\!1)\sigma_{b}^{(v_1(\!e'\!))}
   \sigma_{b}^{(v_2(\!e'\!))}\nonumber
   \\
 \!\!\!\!\!\!  &=& (\!-\!1)^{|E_c|\!+\!\sum_\mu|\!E_\mu\!|}\bigotimes_{\mu}\bigotimes_{e \in E_\mu}
  \sigma_{a_\mu}^{(u(\!e\!))} \sigma_{a_\mu}^{(v(\!e\!))}\nonumber\\
  &&\quad
   \bigotimes_{e' \in E_c}\sigma_{b}^{(v_1(\!e'\!))} \sigma_{b}^{(v_2(\!e'\!))}
   . \nonumber
\end{eqnarray}

We shall take the following convention for $b$ as shown in Table~\ref{tbl:b}. For POVM outcome $a_c=z$, we take
$b=x$; for $a_c=x$, we take $b=z$; for $a_c=y$, we take $b=z$.
With this choice we have
\begin{eqnarray}
  {\cal K}_{{{C}}}&=& (-1)^{|E_c|+\sum_\mu|E_\mu|}\bigotimes_{\mu}(\otimes_{e \in
  E_\mu}\lambda_{u(e)})
  Z_\mu^{|E_\mu|} \nonumber\\
  &&
  \bigotimes_{e\in E_\mu}\sigma_{a_\mu}^{v(e)}\sigma_{b}^{v(e)}
    X_c. \nonumber
\end{eqnarray}
It is convenient to define $n_{\ne b}\equiv \sum_{\mu,a_\mu\ne b}
|E_\mu|$. Then
\begin{eqnarray}
  {\cal K}_{{{C}}}&=& (-1)^{|E_c|+\sum_\mu|E_\mu|}\bigotimes_{\mu}(\otimes_{e \in
  E_\mu}\lambda_{u(e)})
  Z_\mu^{|E_\mu|}\nonumber\\
  && (\bigotimes_{a_\mu\ne b}\otimes_{e \in
  E_\mu}\lambda_{v(e)})Q_c, \label{eqn:stablizer}
\end{eqnarray}
where $Q_c= i^{n_{\ne b}} X_c$ if $n_{\ne b}$ is even and $Q_c= -
i^{1+ n_{\ne b}}(-1)^{\delta_{a_c,x}} Y_c$ if $n_{\ne b}$ is odd.
This gives complete characterization of stabilizer generators, i.e., $Q_c= \pm X_c $ or $Q_c=\pm Y_c$ and the exact sign can be determined.
 This is essential in checking the incompatibility condition.
 
 The above discussion also justifies the assignment of the self-loop associated with the graph staet $|G_0\rangle$:
 \[
 \begin{array}{rl}
n_y \mod 2 =1,& \text{for }T=x,\\
n_x \mod 2 =1, & \text{for }T=y,\\
n_y \mod 2 =1, & \text{for }T=z.
\end{array}\]

Note that the stabilizer operators are not always in the canonical
form in which ${\cal K}_c|_c=X_c$, i.e., they can be $\pm X_c$ or
$\pm Y_c$, but those non-central operators are always $Z$. But it is
easy to find rotations (around logical $z$-axis) to make them
canonical.

 Moreover, as shown in the next Section, in terms
of logical $X$, the measurement operator $O_{c}=(-1)^{|V_c|}
X_c$,  where the $|V_c|$ denotes the number of sites in the domain $c$, so that the effective measurement by $K$ on all sites of a
domain gives rise to a projection $P_c= (1+O_{c})/2$. Taking into account of the probability it corresponds to an operator
$P_c= (1+O_{c})/2^{|V_c|}$ (see below).   If $K|_c=\pm X_c$,
then the induced measurement by $K_\alpha$ corresponds to $X$
measurement in the canonical basis of the standard graph state formulation. If ${\cal K}_c|_c=\pm Y_c$, then
the induced measurement by $K_\alpha$ corresponds to $Y$ measurement
in the canonical basis. Thus whether the induced measurement by
$K_\alpha$ on a domain is in the canonical basis either $X$ or $Y$
can be easily determined by whether $n_{\ne b}$ is even or odd.

\begin{table}
  \begin{tabular}{l|r|r|r}
  $a_c$ & $z$ & $x$ & $y$ \\
  \hline
  $b$ & $x$ & $z$ & $z$\\
  \hline
  $a_{\mu\ne b}$ & $y$ & $y$ & $x$
    \end{tabular}
 \caption{\label{tbl:b} The choice of $b$ and $a_{\mu\ne b}$.}
\end{table}

\section{Proof of the weight formula}

\label{sec:proof2}

Let us mention first the following fact that ($b$ is chosen according to $a_c$ in Table~\ref{tbl:b}),
\begin{equation}
\langle G_0| O_{\rm rest} \otimes_{i\in I_c}\sigma_{b}^{[i]} |G_0\rangle=0,
\end{equation}
if $I_c$ is a strict subset of virtual qubits in any domain ${\cal C}$ (i.e. $|I_c|<4|{\cal C}|$) and $\sigma_b$ is chosen according to Table~\ref{tbl:b} ($O_{\rm rest}$ denotes  operators not in the support of domain ${\cal C}$).
This can easily be proved by the fact that one can choose a stabilizer $S_{jq}\equiv\lambda_j\lambda_q \sigma_{a_c}^{[j]}\sigma_{a_c}^{[q]}$ (see Table~\ref{tbl:encoding}), where $j\in I_c$ and $q\in {\cal C}$ but $q\notin I_c$, so that  $(\otimes_{i\in I_c}(\sigma_{b}^{[i]})$ and $S_{jq}$ anticommutes.  Hence,
\begin{eqnarray*}
&&\langle G_0| O_{\rm rest}  (\otimes_{i\in I_c}\sigma_{b}^{[i]}) |G_0\rangle\\
&=&\langle G_0|O_{\rm rest}  (\otimes_{i\in I_c}\sigma_{b}^{[i]})S_{jk} |G_0\rangle\\
&=&- \langle G_0|S_{jk} O_{\rm rest}  (\otimes_{i\in I_c}\sigma_{b}^{[i]}) |G_0\rangle\\
&=&- \langle G_0| O_{\rm rest}  (\otimes_{i\in I_c}\sigma_{b}^{[i]}) |G_0\rangle,
\end{eqnarray*}
showing that the expectation value is identically zero.

Let us also note the following useful relation regarding to the 4-qubit GHZ associated with the corresponding POVM outcome $K_\alpha$,
\begin{equation}
\label{eqn:usefulGHZ}
 \ketbra{{\rm
GHZ}_\alpha^-}= \Pi_\alpha\frac{(1-\sigma_{b_\alpha}^{[v;1]}\sigma_{b_\alpha}^{[v;2]}\sigma_{b_\alpha}^{[v;3]}\sigma_{b_\alpha}^{[v;4]})}{2}\Pi_\alpha,
 \end{equation}
where $\Pi_\alpha$ ($\alpha=x,y,z$) is a projection to a two-dimensional subspace, equivalently an identity operator on the code subspace  and can be safely omitted when acting on the graph state $|G_0\rangle$. Specificially, $\Pi_x=\ketbra{++++}+\ketbra{----}$, $\Pi_y=\ketbra{i,i,i,i}+\ketbra{-i,-i,-i,-i}$ and $\Pi_z=\ketbra{0000}+\ketbra{1111}$.  The label $b_\alpha$ denotes the corresponding type $b$ if $a_c=\alpha$; see Table~\ref{tbl:b}.

For a given domain (with a given type $\alpha$), the POVM outcome on any site in the domain can be either $F_\alpha $ or $K_\alpha$.  Regarding the number $n_K$ of $K$ outcomes, there are two scenarios: (i)  $n_K$  is less than the total number $|V_c|$ of sites in that domain ${\cal C}$; (ii) $n_K=|V_c|$.

For case (i),  the effect of all those  $K$ in terms of the probability distribution (or the weight formula)  is to multiply a factor of $2^{-n_K}$, i.e., (using $J\in {\cal C}$ to denotes the set of those sites with $K$)
\begin{eqnarray*}
&&\langle G_0| O_{\rm rest} \left(\otimes_{v\in J}  \ketbra{ {\rm GHZ}^{-}_{\alpha(v)}}\right) |G_0\rangle\\
&=&\langle G_0| O_{\rm rest}\otimes_{v\in J} \frac{(1-\sigma_{b}^{[v;1]}\sigma_{b}^{[v;2]}\sigma_{b}^{[v;3]}\sigma_{b}^{[v;4]})}{2} |G_0\rangle\\
&=&2^{-n_K}\langle G_0|O_{\rm rest}   |G_0\rangle,
\end{eqnarray*}
where  $O_{\rm rest}$ denotes  operators not in the support of domain ${\cal C}$, and we have  used
\begin{equation*}
\langle G_0| O_{\rm rest}\otimes_{v\in J} (\sigma_{b}^{[v;1]}\sigma_{b}^{[v;2]}\sigma_{b}^{[v;3]}\sigma_{b}^{[v;4]}) |G_0\rangle=0.
\end{equation*}

For case (ii),   when we expand all the $2^{|V_c|}$ terms in $\otimes_{v\in{\cal C}} (1-\sigma_{b}^{[v;1]}\sigma_{b}^{[v;2]}\sigma_{b}^{[v;3]}\sigma_{b}^{[v;4]})/2$,  the only two nonvanishing contributions are $1/2^{|V_c|}$ and $(-1)^{|V_c|}( \otimes_{i=1}^{4|{\cal C}|}\sigma_{b}^{[i]})/2^{|V_c|}=(-1)^{|V_c|}X_c/2^{|V_c|}$.
In terms
of logical $X$, the effect of all $K$ is equivalent to  $P_c= (1+O_{c})/2^{|V_c|}$, where the  $O_{c}=(-1)^{|V_c|}
X_c$.   That is
\begin{eqnarray*}
&&\langle G_0| O_{\rm rest}\left(\otimes_{v\in C}  \ketbra{ {\rm GHZ}^{-}_{\alpha(v)}}\right)|G_0\rangle\\
&=&\langle G_0| O_{\rm rest}\otimes_{v\in C} \frac{(1-\sigma_{b}^{[v;1]}\sigma_{b}^{[v;2]}\sigma_{b}^{[v;3]}\sigma_{b}^{[v;4]})}{2} |G_0\rangle\\
&=&2^{-|V_c|}\langle G_0|O_{\rm rest} (1+O_c)  |G_0\rangle.
\end{eqnarray*}
Here we also see that the effect of all $K$ in domain ${\cal C}$ is to measurement the logical qubit ${\cal C}$ in the logical $X$, followed by a post-selection of the result corresponding to either positive (if $|V_c|$ is even) or negative (if $|V_c|$ is oddd) eigenvalue of $X$.

With the above preparation, we can move on to the proof. Now consider a spin-2 AKLT state on a bi-colorable lattice ${\cal L}$
(generalization to non-bicolorable lattices is possible), and POVM
elements $F_\alpha$ and $K_\alpha$ ($\alpha=x,y,z$). Denote by $J_F
\subset {\cal L}$ the set of sites where the POVM outcome is of
$F$-type and by $J_K={\cal L}\backslash J_F$ the set of sites where POVM
outcome is of $K$-type.  We should, strictly speaking,
use $\alpha(v)$ to denote the type of $x,y,z$ at site $v$. When
there is no confusion, we simply write $\alpha$.

\smallskip\noindent {\bf Proof of Lemma 1}.  
For simplicity let us denote the AKLT state by $|\psi\rangle$ below. The probability $p(\{F,K\})$ for obtaining POVM measurements $\{F,K\}$ described above is
\begin{eqnarray*}
&& p(\{F,K\})=\langle \psi|\mathop{\otimes}_{v\in J_F}
F_{\alpha(v)}^{(v)\dagger}F_{\alpha(v)}^{(v)}
\mathop{\otimes}_{w\in J_K} K_{\alpha(w)}^{(w)\dagger}K_{\alpha(w)}^{(w)}|{\psi}\rangle\\
&&=\left(\frac{3}{2}\right)^{|J_K|}\langle \psi|\mathop{\otimes}_{v\in
J_F} F_{\alpha(v)}^{(v)\dagger}F_{\alpha(v)}^{(v)} \\
&&\qquad\qquad
\mathop{\otimes}_{w\in J_K} F_{\alpha(w)}^{(w)\dagger} K_{\alpha(w)}^{(w)\dagger}K_{\alpha(w)}^{(w)} F_{\alpha(w)}^{(w)}|{\psi}\rangle\\
&&=\left(\frac{1}{2}\right)^{|J_K|}\langle
\psi|\mathop{\otimes}_{v\in{\cal L}}F_{\alpha(v)}^{(v)\dagger}
\mathop{\otimes}_{w\in J_K} |{\rm
GHZ}_{\alpha(w)}^-\rangle\langle{\rm GHZ}_{\alpha(w)}^-|\\
&&\qquad\qquad
\mathop{\otimes}_{u\in {\cal L}} F_{\alpha(u)}^{(u)} |{\psi}\rangle.
\end{eqnarray*}
In the second equality we have used the fact that $K_\alpha=\sqrt{3/2}K_\alpha F_\alpha$, and in the third equality we have combined all $F$'s
and written explicitly $K_\alpha$'s in terms of the four-qubit GHZ
projectors.

Now we know from Ref.~\cite{WeiAffleckRaussendorf11} that
\begin{equation}
\label{eqn:Fweight2}
\mathop{\otimes}_{u\in {\cal L}}
F_{\alpha(u)}^{(u)} |{\psi}\rangle=c_0\, \left(\frac{1}{\sqrt{2}}\right)^{|{\cal
E}|-|V|}|G_0\rangle,
\end{equation}
where $|G_0\rangle$ is an encoded graph state whose graph $G_0$ is
specified by the POVM elements $\{F\}$, $V$ is the set of domains of
same-outcome POVM measurements, and ${\cal E}$ is the set of
inter-domain edges (before the modulo-2 operation)~\cite{WeiAffleckRaussendorf11}. The
formula~(\ref{eqn:Fweight2}) was originally stated for the honeycomb
lattice, but holds for all bipartite lattices. (For non-bipartite
lattices, an additional condition needs to be imposed relating to
geometric frustration~\cite{Wei13}. Namely, if any domain contains a
cycle with odd number of sites, such $\{F\}$ will not appear.)
Combining the above two results we find that
\begin{eqnarray}
&&p(\{F,K\})=|c_0|^2 \,\left(\frac{1}{2}\right)^{|{\cal E}|-|V|+ |J_K|}\nonumber\\
&&\quad\times\langle G_0|\left(\mathop{\otimes}_{v\in J_K} |{\rm
GHZ}_{\alpha(v)}^-\rangle\langle{\rm GHZ}_{\alpha(v)}^-|  \right)
|G_0\rangle.
\end{eqnarray}

Using Eq.~(\ref{eqn:usefulGHZ}) and the results in the beginning of the section, we know that for those GHZ-projections in a domain such that their number is less than the total number of sites in the domain, i.e., case (i) discussed above, their contribution is to mulitiply by a factor $2^{-n_K}$. For those such that the two numbers are equal, i.e., case (ii), these GHZ-projections (in a domain) can be replaced by $P_c= (1+O_{c})/2^{|V_c|}$, where the  $O_{c}=(-1)^{|V_c|}
X_c$, and $c$ labels the domain. 
Thus,
\begin{eqnarray}
&&p(\{F,K\})=|c_0|^2 \,\left(\frac{1}{2}\right)^{|{\cal E}|-|V|+ 2|J_K|}\nonumber\\
&&\quad\times\langle G_0|\mathop{\otimes}_{c\in D_K}(I_c+O_{c}) 
|G_0\rangle,
\end{eqnarray}
where we use $D_K$ to label the domains that contain the same number of $K$ operators as the total number of internal sites. 

Next we demonstrate the first part of the Lemma. Assume that,
for some subset $Q\in D_K$, the observable $
-\mathop{\otimes}_{c\in Q}O_c \in {\cal S}(|G_0\rangle)$. Then,
\begin{eqnarray}
&&\langle G_0|\mathop{\otimes}_{\mu\in D_K}({I_\mu+ O_\mu})|G_0\rangle\nonumber\\
&=&\langle G_0|\mathop{\otimes}_{\nu\in D_K\backslash Q}({I_\nu+ O_\nu})\mathop{\otimes}_{\mu\in Q}({I_\mu+ O_\mu})\left(-\mathop{\otimes}_{c\in Q}O_c\right)|G_0\rangle\nonumber\\
&=&-\langle G_0|\mathop{\otimes}_{\nu\in D_K\backslash Q}({I_\nu+ O_\nu})\mathop{\otimes}_{c\in Q}({O_c+ I_c})|G_0\rangle\nonumber\\
&=&-\langle G_0|\mathop{\otimes}_{\mu\in D_K}({I_\mu+ O_\mu})|G_0\rangle=0.\nonumber
\end{eqnarray}
In the third line we have used the fact $O_\mu^2=I_\mu$. Let us also note that being product of Pauli operators, $O_\mu$ either commutes or anticommutes with another product of Pauli operators.

Next, we demonstrate the second part of the Lemma, i.e., finding $p(\{F,K\})$ when it is not identically zero.
 Consider a subset of domains $Q\subset D_K$. If $\mathop{\otimes}_{\mu\in Q} O_\mu \not\in \pm{\cal S}(|G_0\rangle)$, then $\langle G_0|\mathop{\otimes}_{\mu\in Q} O_\mu|G_0\rangle=0$ (note that $\mu$ is an index for the domain, not an index for the site). Furthermore, if the incompatibility condition is not satisfied, then  $\mathop{\otimes}_{\mu\in {Q}}O_\mu \in \pm {\cal S}(|G_0\rangle)$ implies that
$\mathop{\otimes}_{\mu\in {Q}}O_\mu \in  {\cal S}(|G_0\rangle)$, and therefore  $\langle G_0|\mathop{\otimes}_{\mu\in {Q}}O_\mu |G_0\rangle=1$. We now exapnd the projector $\mathop{\otimes}_{c\in D_K} (I_c +O_c)$ in the matrix element,
\begin{eqnarray}
&&\langle G_0|\mathop{\otimes}_{c\in D_K} (I_c+O_c) |G_0\rangle\\
&=& \langle G_0|\sum_{Q\subset D_K} \mathop{\otimes}_{\mu\in Q} O_\mu|G_0\rangle ={|M|},
\end{eqnarray}
where the set $M$ is defined as $M=\{O(Q)\equiv \mathop{\otimes}_{\mu\in Q} O_w | Q\subset D_K\, \mbox{and}\, O(Q)\in {\cal S}(|G_0\rangle)\}$. Actually $M$ has the following equivalent formulation which will turn out to be useful,
\begin{eqnarray}
M&=&\{O(Q)\equiv \mathop{\otimes}_{\mu\in Q} O_\mu | Q\subset D_K\, \nonumber\\
&&\mbox{and}\,  [O(Q),S]=0, \forall S\in  {\cal S}(|G_0\rangle)\}.
\end{eqnarray}
Using this latter characterization of $M$, we now turn to the counting for $|M|$. We describe every subset $Q$ of $D_K$ by its characteristic vector ${\bf q}$, defined as follows:  if $\mu\in Q$ then $q_\mu=1$, or if $\mu\not\in Q$, then $q_\mu=0$. Furthermore we define a binary-valued matrix $H$ of dimension $|{ V}|\times |D_K|$ (where $|{ V}|$ denotes total number of domains), whose entries are
\begin{eqnarray}
&&H_{\mu\nu}=0, \ \mbox{if} \ [{\cal K}_\mu, O_\nu]=0, \nonumber\\
&&H_{\mu\nu}=1, \ \mbox{if} \ \{{\cal K}_\mu, O_\nu\}=0, \nonumber
\end{eqnarray}
where $\mu \in {V}$ (the set of all domains) and $\nu\in D_K$ (the set of those domains with equal number of $K$'s and sites). Then for any $Q\subset D_K$, $O(Q)\in M$ if and only if $H {\bf q} \,{\rm mod}\, 2 ={\bf 0}$. Therefore,
\begin{equation}
|M|=2^{\dK}.
\end{equation}
Putting everything into the expression for $p(\{F,K\})$ we obtain the equation~(\ref{eqn:pWeight}),
\begin{equation*}
p(\{F,K\})=|c_0|^2 \,\left(\frac{1}{2}\right)^{|{\cal E}|-|V|+2 |J_K|-\dK},
\end{equation*}
  and the Lemma is proved.

We remark that checking the kernel of a binary matrix can be done via, e.g., the Gauss elimination method; see e.g.~\cite{KocArachchige}. Furthermore, to check the incompatibility condition it is sufficient to check the products of $O_\mu$ associated with all basis vectors ${\bf q}$'s in the kernel. If none of them satisifies it, then the incompatibility condition is not satisified.

\end{document}